\documentclass[titlepage,12pt]{article}
\usepackage{textcomp}
\usepackage{amssymb,amsmath,color,graphics,amscd,epsf,indentfirst,amsfonts}
\usepackage{mathtools}
\usepackage{enumerate}
\usepackage{epsfig}
\usepackage{slashed}
\usepackage{comment}

\usepackage{parskip}

\usepackage{hyperref}
\hypersetup{
    colorlinks=true,
    linkcolor=blue,
    filecolor=magenta,      
    urlcolor=cyan
    }

\usepackage[margin=1in,footskip=0.5in]{geometry}

\usepackage{tikz}
\usetikzlibrary {arrows.meta}
\usetikzlibrary{decorations.pathmorphing}
\tikzset{snake it/.style={decorate, decoration=snake}}

\def\blfootnote{\xdef\@thefnmark{}\@footnotetext}

\long\def\symbolfootnote[#1]#2{\begingroup%
\def\thefootnote{\fnsymbol{footnote}}\footnote[#1]{#2}\endgroup}

\makeatletter
\renewcommand{\@dotsep}{4.5}
\makeatother

\def\be{\begin{equation}}
\def\ee{\end{equation}}

\makeatletter
\def\@seccntformat#1{\csname the#1\endcsname.\quad}
\makeatother

\def\clock{{\count0=\time
           \divide\count0 60
           \ifnum\count0<10 0\fi\the\count0
           \multiply\count0 -60 \advance\count0 \time
           :\ifnum\count0<10 0\fi \the\count0
         }}
\newcommand{\timestamp}{{\small\vbox{\hbox{\tt\jobname.tex}
\hbox{\the\day/\the\month/\the\year, \clock}}}}

\def\time{{\tau}}

\def\beq{\begin{equation}}
\def\eeq{\end{equation}}
\newcommand{\bea}{\begin{eqnarray}}
\newcommand{\eea}{\end{eqnarray}}
\def\bal{\begin{align}}
\def\eal{\end{align}}

\def\drawbox#1#2{\hrule height#2pt
         \hbox{\vrule width#2pt height#1pt \kern#1pt
               \vrule width#2pt}
               \hrule height#2pt}

\def\Asym#1#2{\vcenter{\vbox{\drawbox{#1}{#2}
               \kern-#2pt       
               \drawbox{#1}{#2}}}}

\newcommand{\qq}{\quad\quad}

\numberwithin{equation}{section}

\begin{document}
\begin{titlepage} 
\vskip 4cm
\begin{center}
\font\titlerm=cmr10 scaled\magstep4
    \font\titlei=cmmi10 scaled\magstep4
    \font\titleis=cmmi7 scaled\magstep4
    \centerline{\LARGE \titlerm 
      Domain Walls in Large-$N$ QCD$_3$}
          \vskip 0.3cm
\vskip 1cm
       {Adi Armoni, Jonathan Whittle}\\
\vskip 0.5cm
       {\it Department of Physics, Faculty of Science and Engineering}\\
       {\it Swansea University, SA2 8PP, UK}\\       

\vskip 0.5cm
{a.armoni@swansea.ac.uk, jonathan.whittle@swansea.ac.uk}\\

\end{center}
\vskip .5cm
\centerline{\bf Abstract}
\baselineskip 20pt
%

\vskip .5cm 
\noindent

Large-$N$ three dimensional QCD with a level $k$ Chern-Simons term and $N_f$ massless flavours admits $N_f+1$ degenerate vacua. We discuss various aspects of the domain walls that interpolate between those vacua. Using a bosonic dual for the case $k \ge N_f/2$ we write down the profile of the composite wall. By fluctuating the profile, we find the 2d field theory that lives on the wall, a level-$N$ gauged WZW model, and compare it with a theory that was previously obtained by using anomaly inflow considerations. We also discuss the case $k < N_f/2$. Finally, we remark on the realisation of the wall as a D-brane in a holographic dual.

\vfill
\noindent
\end{titlepage}\vfill\eject

\setcounter{equation}{0}

\pagestyle{empty}
\small
\normalsize
\pagestyle{plain}
\setcounter{page}{1}

\tableofcontents
 
\section{Introduction} 
\label{intro}

    Domain walls, extended objects that interpolate between multiple vacua of a given theory, play a  central role in various fields of physics, including condensed matter, high energy, strings and cosmology. Here we focus on field theory and string theory.
    
    This paper concerns the vacua of three dimensional large-$N$ QCD. We consider a 3d $SU(N)$ gauge theory coupled to $N_f$ fermions in the fundamental representation and a level $k$ Chern-Simons term. It was shown in \cite{armoni_2020_metastable} that in the 't Hooft large-$N$ limit the theory admits $N_f+1$ degenerate vacua. The degeneracy is lifted at finite $N$. We therefore anticipate the existence of domain walls whose tension is ${\cal O}(N)$ and their lifetime is $T \sim \exp N$.
    
    Domain walls can be studied from various angles, in particular as a classical profile that interpolates between distinct vacua, namely a field configuration whose asymptotic values corresponds to different vacua of a given theory. We begin by carefully studying the profile of the fundamental domain wall, following \cite{jackiw_1990_selfdual}, namely the wall that interpolates between neighbouring vacua. We write down the profile of both the gauge field and the scalar that corresponds to the bosonised fermion. By fluctuating the wall we find the two dimensional field theory that lives on the wall. In the case of the the fundamental wall it consists of two scalars: a 'centre of mass' field and a scalar that corresponds to the fluctuation of the bosonised field.  

    We then generalise our discussion to the composite (non-abelian) wall. We begin with the $k \ge N_f/2$ case, where the QCD theory is described by a $U(N_f/2+k)$ bosonic dual with a level $-N$ Chern-Simons term. By considering the non-abelian profile and its fluctuations, we find a level $N$ gauged WZW theory accompanied by a 'centre of mass' scalar on the wall. In addition we also study the case $k < N_f/2$ where the QCD theory is described by two bosonic duals, each of which covers half of the vacua.
    
    From the profile, we can read the tension of the wall and compare it to our expectation. Specifically, in the current setup we expect a wall with a tension $N$ \cite{armoni_2020_metastable}, a very heavy object. 
    
    In addition to the above we can partially deduce the field theory on the wall by using anomaly considerations \cite{Armoni:2015jsa}: since the distinct vacua of the theory admit different Chern-Simons theories, there is apriori a gauge anomaly localised on the wall. The wall has to absorb the anomaly deficit to ensure that the full configuration is anomaly free. A similar conclusion arises from the holographic dual \cite{argurio_2020_vacuum}, where the vacua of the QCD theory are described in terms of a distribution of 'flavour branes' and a 'Chern-Simons brane'. 
    In all string setups the wall is represented by a D-brane. The tension of a D-brane is $\sim 1/g_{\rm st}\sim N$, hence we find an agreement with field theory.  

    The massless modes on a composite wall admit a $U(n)$ global symmetry\footnote{To be precise, this comment is valid for $k>N_f/2$. We expect a similar picture for $k\le N_f/2$.}, where $n$ is the 'distance' between the vacua ($n=p'-p$, with $p,p'=0,1,...,N_f$ labelling the vacua). The wall can be viewed as a bound state of $n$ fundamental walls. In the large-$N$ limit the dynamics on the wall is dominated by the Cartan, namely a $U(1)^n$ theory with wall tension $\sim nN$. This is in agreement with the expectation that in the large-$N$ limit the composite wall is reduced to $n$ fundamental walls, whose couplings are $\sim 1/N$. In the large-$N$ limit the wall is described by $n$ real massless scalars, whose expectation values parametrise the positions of the wall's constituents. There is a $1/N$ attractive potential between the $n$ scalars, such that at finite $N$ only the centre of mass remains massless and the rest become massive and can be integrated out.
    
    The paper is organised as follows: in section \eqref{back} we review large-$N$ QCD\textsubscript{3} and its bosonic dual. In section \eqref{prof} we discuss the wall profile, focusing on the fundamental wall.  Section \eqref{sec: field} is devoted to the study of the 2d field theory that lives on the fundamental wall. In sections \eqref{sec: composite wall} and \eqref{sec: composite wall FT} we generalise our discussion to the case of a composite (non-abelian) wall. In section \eqref{sec:two_bosonic} we discuss the domain wall solutions when $k<N_f/2$, and in section \eqref{sec: D-brane} we comment on the realisation of the wall as a D-brane. We conclude in section \eqref{sec: conc}.

\section{Background} 
\label{back}

    We briefly summarise the content of \cite{armoni_2020_metastable}.
    Consider three dimensional $SU(N)$ Yang-Mills theory coupled to $N_f$ fermion flavours $\psi_a^i$ that transform in the fundamental representation. In addition let us add a level $k$ Chern-Simons term, so the action is:
    
    \begin{equation}\label{eqn: fermionic theory}
        S = \int d^3x \ \mathrm{tr} \left[ \frac{1}{g^2} F_{\mu\nu} F^{\mu\nu} + \frac{k}{4\pi} \epsilon^{\mu\nu\rho} ( C_\mu \partial_\nu C_\rho + \frac{2i}{3} C_\mu C_\nu C_\rho) + i \bar{\psi}_i \slashed{D} \psi^i - m_i^j \bar{\psi}^i \psi_j   \right]
    \end{equation}
    
    where the trace is taken over gauge indices. We focus on the limit $m_i ^j \rightarrow 0^+$. In the large-$N$ limit ($N\rightarrow \infty$ with $\Lambda = g^2 N$ fixed), with fixed $N_f$ and fixed $k$ the theory can be analysed using the method of Coleman and Witten \cite{Coleman:1980mx}. We can introduce a meson field $M_i^j = \bar{\psi}_i^a \psi^j_a / N$ and determine the properties of its potential by considering the disk diagram which consists of a single fermion loop filled with planar gluon propagators. These diagrams are order $N$ and dominate in the large-N limit.

    We can diagonalise the meson using a $U(N_f)$ flavour transformation. Unlike the four dimensional case where in the vacuum the eigenvalues of the meson field are identical and the meson field expectation value is proportional to the $N_f \times N_f$ unit matrix, the eigenvalues of the meson field in the three dimensional theory can be independently either $+1$ or $-1$, due to parity. To be more concrete, in each vacuum the meson takes the form
    
    \begin{equation}
        M=\rm diag (\lambda_1, \lambda _2, ... , \lambda_{N_f}) \, ,
    \end{equation}
    
    with $\lambda_i =\pm 1$. A generic vacuum consists of $p$ eigenvalues with $\lambda_i=+1$ and $N_f-p$ eigenvalues with $\lambda_i=-1$. Therefore, there are $N_f+1$ vacua, each labelled by an integer $p \in (0, 1, ... , N_f)$. The $U(N_f)$ flavour symmetry breaking pattern in a given vacuum is
    
    \begin{equation}
    U(N_f)\rightarrow U(p) \times U(N_f-p)
    \end{equation}
    
    In addition, in each vacuum there is a Chern-Simons term with shifted level $k \rightarrow k'=k+p-N_f/2$. This means that the vacuum theory in each of the $N_f+1$ sectors is:

    \begin{equation}
        SU(N)_{k+p-N_f/2} \otimes \mathrm{Gr}(N_f,p)
    \end{equation}
    
    where Gr$(N_f,p)$ denotes the Grassmannian $U(N_f)/(U(p)\times U(N_f-p))$ from the flavour symmetry breaking, which is the target space for a Goldstone boson NLSM in each sector. The $SU(N)_{k'}$ Chern-Simons theory is level-rank dual to a $U(k')_{-N}$ Chern-Simons theory, which allows for the construction of the bosonic dual theory discussed below.

    Taking nonzero masses for the fermions, or going beyond leading order in the $1/N$ expansion lifts the vacuum degeneracy and singles out one sector as the true vacuum.
    
    The vacua of the QCD theory can be equivalently described by a bosonic dual. When $k \ge N_f/2$ the bosonic theory consists of $N_f$ fundamental scalars coupled to a $U(\tilde N)$ gauge field, where $\tilde N = N_f/2 +k$ and there is a Chern-Simons term with level $\tilde{k} = -N$. In addition, there is a sextic potential for the scalars added by hand in order to reproduce the desired set of $N_f + 1$ vacua.
    
    The action of the bosonic dual is given by
    
    \begin{equation}\label{eqn: bosonic dual}
        S = N \int d^3 x \ \mathrm{tr} \Big( (D_\mu \Phi) (D^\mu \Phi)^\dagger + \frac{1}{4 \pi} \epsilon^{\mu\nu\rho}( A_\mu \partial_\nu A_\rho + \frac{2}{3} A_\mu A_\nu A_\rho) - \Lambda^3 | \Phi |^2 (|\Phi|^2 -u^2 )^2 \Big)
    \end{equation}
    
    where $\Lambda = g^2 N$ is the QCD mass scale, the trace is over gauge indices, and we have rescaled the scalar field $\Phi$ such that $N$ multiplies the entire action. The meson field of the scalar theory $\tilde{M}=\Phi ^\dagger \Phi$ is identified (up to a rescaling + shift) with the meson field of the original QCD theory.
    The vacua of the theory are obtained by locating the scalars at the minima of the potential at either $\Phi=0$ or $\Phi=u$. When we have $p$ scalars in the $0$ minimum and $N_f - p$ in the $u$ minimum the gauge group is Higgsed to $U(\tilde{N}) \rightarrow U(k+p-N_f/2)$, and the flavour symmetry is broken as $U(N_f) \rightarrow U(p) \times U(N_f-p)$. The resulting vacuum theory is:

    \begin{equation}\label{eqn: bosonic dual vacua}
        U(k+p-N_f/2)_{-N}  \otimes \mathrm{Gr}(N_f,p)
    \end{equation}
    
    which is level-rank dual to the corresponding vacuum theory of the original QCD theory. 

    When $k<N_f/2$ the vacua of QCD\textsubscript{3} cannot be obtained from a single bosonic dual theory - instead a pair a bosonic duals must be employed: a $U(N_f/2-k)_{-N}$ theory and a $U(N_f/2+k)_N$ theory. The $N_f$ scalar fields $\Phi$ are this time charged under both gauge groups, and after condensation each of the low-energy theories in \ref{eqn: bosonic dual vacua} can be obtained via Higgsing and symmetry breaking.
    
    There is another useful description of the vacua of the theory, obtained by holography \cite{argurio_2020_vacuum}. The description incorporates both the QCD theory and its bosonic duals for all $k$.

\section{Wall Profile} 
\label{prof}

    In this section we construct the domain wall profile. We focus on the 'fundamental wall' which in the bosonic dual description is obtained by considering an Abelian gauge theory with a single scalar field flavour. The bosonic meson has a single eigenvalue $\lambda$, and the wall interpolates between the vacuum with $\lambda=0$ and the vacuum with $\lambda=u^2$. 

    We re-derive the results of \cite{jackiw_1990_selfdual} where an Abelian Chern-Simons Higgs model is analysed, and vortex and domain wall solutions are found. Other works concerning solitonic solutions to the Abelian Chern-Simons Higgs model include \cite{Kao:1996tv, Edelstein:1993bb}. There is a $U(1)$ gauge field $A_\mu$, a single complex scalar $\Phi$ with charge $e$ which we set to $1$, and a sextic potential for $\Phi$ which is renormalisable in $d=2+1$ and is compatible with $\mathcal{N}=2$ supersymmetry. In this section we derive and solve the domain wall equations for this system, using a slightly different approach to that in \cite{jackiw_1990_selfdual}. This will give us the domain wall which interpolates between neighbouring vacua of QCD\textsubscript{3}, i.e. between vacua labelled by $p$ and $p+1$.

    In order to obtain this model as the bosonic dual of QCD\textsubscript{3} we need to start from an $SU(N)_{1/2}$ Chern-Simons theory with $N_f=1$ flavours of fermion. Then, following the proposal of \cite{armoni_2020_metastable} leads to the following action for the bosonic dual:

    \begin{equation}\label{eqn: dual_lagr}
        S = N \int d^3x \ \left[ (D_\mu \Phi)^* D^\mu \Phi + \frac{1}{4\pi} \epsilon^{\mu\nu\rho} A_\mu \partial_\nu A_\rho - \Lambda^3 |\Phi|^2 (|\Phi|^2 - u^2)^2 \right]
    \end{equation}

    where the covariant derivative is $D_\mu \Phi = \partial_\mu \Phi + iA_\mu \Phi$, and $\Lambda$ is the QCD mass scale from the original theory \eqref{eqn: fermionic theory}. The fields transform under a gauge transformation like:

    \begin{equation}
        \Phi \rightarrow \Phi e^{-i\lambda} \qq \qq A_\mu \rightarrow A_\mu + \partial_\mu \lambda.
    \end{equation}

    We can split $\Phi$ up into two real-valued fields $\rho$ and $\varphi$ like $\Phi = \rho e^{i\varphi}$ which means that $\rho$ is gauge-invariant but $\varphi$ transforms like $\varphi \rightarrow \varphi - \lambda$. We can then form the gauge-invariant combination $A'_\mu = A_\mu +  \partial_\mu \varphi$ and rewrite the action as:
    
    \begin{equation}\label{eqn: dual_lagr2}
        S = N \int d^3x \ \left[ \partial_\mu \rho \partial^\mu \rho + \rho^2 A'_\mu A'^\mu + \frac{1}{4\pi} \epsilon^{\mu\nu\rho} A'_\mu \partial_\nu A'_\rho - \Lambda^3 \rho^2 (\rho^2 - u^2)^2 \right]
    \end{equation}

    where we have discarded a $- \frac{N}{4\pi} \epsilon^{\mu\nu\rho} \partial_\mu \varphi \partial_\nu A_\rho = - \partial_\mu( \frac{N}{4\pi} \epsilon^{\mu\nu\rho} \varphi \partial_\nu A_\rho)$ boundary term. The equations of motion associated with the fields in this action are:

    \begin{equation}
        \partial_\mu \partial^\mu \rho - \rho A'_\mu A'^\mu = - \frac{1}{2} \frac{\partial V}{\partial \rho} \qq \qq \frac{1}{4\pi} \epsilon^{\mu\nu\rho} F'_{\nu\rho} = - 2 \rho^2 A'^\mu.
    \end{equation}

    From the second equation we can define the conserved (topological) $U(1)$ current $J^\mu = \epsilon^{\mu\nu\rho} F'_{\nu\rho}/4\pi$ which obeys $\partial_\mu J^\mu = 0$ by virtue of the antisymmetric tensor. This kind of current is a feature common to $U(N)$ Chern-Simons theories. The time component relates electric and magnetic charges:

    \begin{equation}\label{eqn: csglaw}
        \frac{1}{4\pi} B = -\frac{1}{2} J^t \qq \qq \implies \qq \qq A'^t = \frac{B}{4\pi \rho^2}
    \end{equation}

    where the magnetic field $B = -F_{xy}$, and the electric charge is obtained from $Q = \int d^2 \boldsymbol{x} \ J^t$. This means that any object with magnetic charge also carries electric charge and is therefore a dyon (this is characteristic of Chern-Simons theories). The Chern-Simons level $N$ doesn't appear in the above relation, which is unusual, however this is just a consequence of our $\rho$ scaling. To recover the usual relation we can rescale $\rho \rightarrow \rho/\sqrt{N}$ and we instead end up with $NB/4\pi = -J^t/2$.

    The energy-momentum tensor is found by varying the action with respect to the metric:

    \begin{equation}
        T_{\mu\nu} = \frac{2}{\sqrt{-g}} \frac{\delta S}{\delta g^{\mu\nu}} \qq \implies \qq T_{\mu\nu} = 2N (D_\mu \Phi)^*D_\nu\Phi - N g_{\mu\nu}\Big[ (D^\alpha \Phi)^*D_\alpha\Phi - \Lambda^3 |\Phi|^2(|\Phi|^2 - u^2)^2 \Big]
    \end{equation}

    where we raise and lower indices using $g_{\mu\nu} = \mathrm{diag}(-1,+1,+1)$. The Chern-Simons term is absent because it is topological and does not depend on the metric. Now, employing equation \eqref{eqn: csglaw}, we can write the energy as:

    \begin{equation}
        E = T_{tt} = N \int d^2 \boldsymbol{x} \ \Big[ (\partial_t \rho)^2 + (\partial_x \rho)^2 + (\partial_y \rho)^2 + \frac{B^2}{16 \pi^2 \rho^2} +  \rho^2 A'_xA'_x +  \rho^2 A'_yA'_y + \Lambda^3 \rho^2 (\rho^2 - u^2)^2 \Big]. 
    \end{equation}

    From this we can read off the boundary conditions that finite-energy field configurations must obey. Firstly all $\rho$ derivatives must vanish, and then from the potential we see that either $\rho=0$ or $\rho = u$ at spatial infinity (the scalar field must be in one of its vacua). The $A'_x$ and $A'_y$ 'gauge' field components must be zero when $\rho$ is in the symmetry-breaking vacuum ($\rho=u$), but are unconstrained when $\rho$ is in the symmetric vacuum ($\rho=0$). In order for the magnetic field term to vanish at spatial infinity, we also require $\partial_x A'_y=0$ at infinity.

    In order to look for time-independent domain wall solutions parallel to the $y$ axis, we set $\partial_t = \partial_y = 0$ and the energy admits the Bogomol'nyi completion:

    \begin{align}
        E & = N \int d^2 \boldsymbol{x} \ \Big[ \left (\partial_x \rho \mp \Lambda^{3/2} \rho(\rho^2-u^2) \right )^2 + \left ( \rho A'_y \mp \frac{\partial_x A'_y}{4 \pi  \rho} \right )^2 +  \rho^2 A'_x A'_x \Big] \pm
         \nonumber \\ &
        N L_y \Big[ \frac{\Lambda^{3/2}}{2}(\rho^2-u^2)^2 + \frac{1}{4\pi} A'^2_y \Big]^{x=+\infty}_{x=-\infty}  \geq \pm N L_y \Big[ \frac{\Lambda^{3/2}}{2}(\rho^2-u^2)^2 + \frac{1}{4\pi} A'^2_y \Big]^{x=+\infty}_{x=-\infty}
    \end{align}

    where we have used that $B = -\partial_xA'_y$ for our $y$-independent ansatz. Performing the integral over the $y$ direction gives us the volume factor $L_y$. Configurations possessing the minimum energy saturate the inequality (the Bogomol'nyi bound) and must satisfy the following three Bogomol'nyi equations:

    \begin{align}
        \partial_x \rho \mp \Lambda^{3/2} \rho(\rho^2-u^2) = 0 \\  \rho^2 A'_y \mp \frac{1}{4 \pi} \partial_x A'_y =0 \\ \rho^2 A'_x A'_x=0
    \end{align}

    which are solved by:

    \begin{equation}\label{eqn: fund solns}
        \rho = u(1+e^{2u^2 \Lambda^{3/2}(x+X)})^{-1/2} \,\,\, , \,\,\,  A'_y = j (1+e^{-2u^2 \Lambda^{3/2}(x+X)})^{-2\pi /\Lambda^{3/2}} \,\,\, , \,\,\, A'_x = 0
    \end{equation}

    where $X$ is the centre of mass position of the wall and $j$ is unconstrained, according to the boundary conditions described above. We have taken the upper signs in the factorisation of the energy. Perturbing these solutions will take us away from the Bogomol'nyi limit which necessarily increases the energy, and therefore the domain wall is stable against decay. (The exception is if we translate the wall in the $x$ direction - there is no change in the energy when we do this which means that $X$ is a collective coordinate of the domain wall solution.)
    
\begin{figure}
    	\centering
    	\includegraphics[width = 5in]{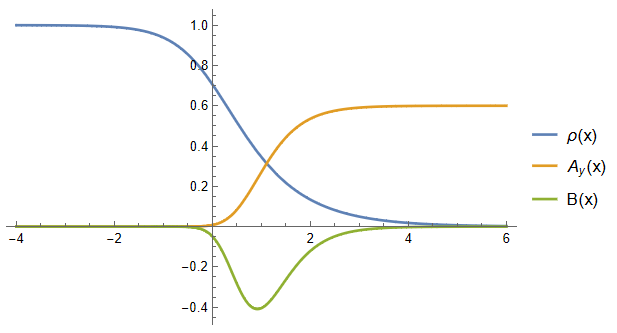}
    	\caption{The form of the domain wall solution in equation \eqref{eqn: fund solns} for the (arbitrary) choice $j=0.6$, and $u=1$, with scalar field in blue, 'gauge' field in orange, and magnetic field in green. The wall centre of mass $X$ has been chosen to be zero, but changing it would translate all three components together.}
   	\label{fig: domain wall}
    \end{figure}

    There is a magnetic field localised at the wall which we can obtain using $B = - \partial_x A'_y$ and if we integrate this over $x$ we find that $-j$ is the magnetic flux per unit length carried by the wall. From equation \eqref{eqn: csglaw} we can see that the wall also carries electric charge $Q = j/2\pi$. The domain wall profile is shown in figure \eqref{fig: domain wall}.

    The tension per unit length of the wall is given by the Bogomol'nyi bound, with the appropriate boundary conditions at $x=\pm\infty$ applied. It reads:

    \begin{equation}\label{eqn: abelian tension}
        T = N \Big[ \frac{u^4 \Lambda^{3/2}}{2} + \frac{1}{4\pi}j^2 \Big]
    \end{equation}

    so there are contributions from both scalar and 'gauge' parts, both of which are $\mathcal{O}(1)$ in the large-N limit. The tension goes as $N$, which is to be expected since the QCD\textsubscript{3} vacua are separated by energy barriers with a height of order $N$.

    We can rewrite the tension in terms of the electric charge $Q$ rather than the magnetic flux $j$. After performing a canonical rescaling which takes $\rho \rightarrow \rho / \sqrt{N}$, the parameters are related by $j = 2 \pi Q / N$ which means that the tension can be written as:

    \begin{equation}
        T = N \Big[ \frac{u^4 \Lambda^{3/2}}{2} + \frac{\pi}{N^2}Q^2 \Big]
    \end{equation}

    so if we fix the charge to be $\mathcal{O}(1)$ then the magnetic flux is $1/N$ suppressed and the gauge field contribution is suppressed by an overall factor of $1/N$.

    The $A'_t$ 'gauge' field component is not determined by the Bogomol'nyi equations but we can use the equation of motion \eqref{eqn: csglaw} to find it in terms of $\rho$ and $A'_y$. We find that it also has an interpolating form given by:

    \begin{equation}
        A'^t = - j (1+e^{-2u^2 \Lambda^{3/2}(x+X)})^{-2\pi /\Lambda^{3/2}}
    \end{equation}

    so after lowering the index, we find that $A'_t = A'_y$. Note that this $j$ is the same (arbitrary) constant as that appearing in the $A'_y$ profile.

    We can calculate the domain wall momentum from the other components of the energy-momentum tensor. We find that $P_x = T_{tx}$ vanishes, but the momentum density in the $y$ direction is given by:

    \begin{equation}
        P_y = T_{ty} = 2N \int dx \ \rho^2 A'_t A'_y = \frac{Nj^2}{4\pi}
    \end{equation}

    Since the domain wall is extended in the infinite $y$ direction this doesn't violate our static solution ansatz. In terms of the electric charge, the momentum reads (after canonical rescaling) $P_y = \pi Q^2 / N$ so it is suppressed by a factor of $1/N$ if we fix the charge to be $\mathcal{O}(1)$.

\section{Field theory on the fundamental wall} 
\label{sec: field}

    We would like to determine the $1+1$ dimensional field theory which lives on the domain wall world-volume for our Abelian, single-flavour wall. There are a variety of different methods available for deducing the modes living on the wall \cite{ Armoni:2015jsa, Acharya:2001dz, Auzzi:2008zd, Shifman:2002jm, Bashmakov:2018ghn, Davidson:2002eu, Dvali:2007nm, Garriga:1991ts, Armoni:2005sp,Gaiotto:2017tne}, some of which employ holographic models like those we will describe in section \eqref{sec: D-brane}. We will start from the action after field redefinitions:

    \begin{equation}
        S = N \int d^3x \ \left[ \partial_\mu \rho \partial^\mu \rho + \rho^2 A'_\mu A'^\mu + \frac{1}{4\pi} \epsilon^{\mu\nu\rho} A'_\mu \partial_\nu A'_\rho - \Lambda^3 \rho^2 (\rho^2 - u^2)^2 \right].
    \end{equation}

    Now that we have found a domain wall solution, the boundary term which we discarded becomes important, since the wall acts as an interface localised at $x=-X$ between two different theories. To obtain the theory on the wall, we take the solutions \eqref{eqn: fund solns} that we found in the previous section and then we upgrade the two parameters $X$ and $j$ to fields\footnote{What we are actually doing is fluctuating $j$ around a fixed value which we set (arbitrarily) after solving the Bogomol'nyi equation, so we should more properly write $j \rightarrow j + \delta_j (t,y)$ and then our 2d action would contain terms for $\delta_j$. This is equivalent however to taking $j \rightarrow j(t,y)$ which is what we do in this section, although while the Bogomol'nyi equations are solved with $j_t = j_y$, we need to remember that this no longer holds if we are fluctuating the $j$s separately.} depending on the world-volume coordinates $(t,y)$, and substitute these into the 3d action and perform the integral over $x$. We find that the action reduces to two dimensions like:

    \begin{equation}\label{eqn: 2daction}
        S_{2d} = N\int d^2x \ \Big[ \frac{u^4 \Lambda^{3/2}}{4} \partial_m X \partial^m X + \frac{1}{8\pi} j_m j^m - \frac{1}{8\pi} \epsilon^{mn} \varphi \mathcal{J}_{mn} \Big]
    \end{equation}

    where $\mathcal{J}_{mn}= \partial_m j_n - \partial_n j_m$ is the field strength for the 2d gauge field $j_m$, and we have defined the 2d epsilon tensor like $\epsilon^{mn} = \epsilon^{mxn}$ so that $\epsilon^{ty}=+1$ and $\epsilon^{yt}=-1$. We have also ignored a constant term.

    Chern-Simons theories defined on manifolds with boundaries are not gauge-invariant. Our domain wall acts like a boundary located at $x=-X$, between low-energy theories which in this case are a $U(1)_N$ pure Chern-Simons theory on one side, and nothing (i.e. a gauge theory with fully Higgsed gauge group) on the other. A gauge transformation therefore acts on the bulk action like:

    \begin{equation}
        S_{3d} = \frac{N}{4\pi} \int_{x > -X} d^3 x \ \epsilon^{\mu\nu\rho} A_\mu \partial_\nu A_\rho \,\,\, \rightarrow \,\,\, \frac{N}{4\pi} \int_{x > -X} d^3 x \ \epsilon^{\mu\nu\rho} A_\mu \partial_\nu A_\rho - \frac{N}{4\pi} \int_{x=-X} d^2 x \ \epsilon^{mn} \lambda \partial_m A_n
    \end{equation}

    and it is necessary to add suitable boundary terms which by their own gauge transformations cancel this one. The term which achieves this is the final term in \eqref{eqn: 2daction}, which means that the combined bulk plus boundary action is gauge invariant.

    We can write the 2d action in a different form by undoing our earlier field redefinition. In the same way that we switched from the gauge field $A_\mu$ to a gauge-invariant field $A'_\mu$ here we will switch from the gauge-invariant field $j_m$ to a $U(1)$ gauge field $a_m$ using

    \begin{equation}
        a_m = j_m - \partial_m \varphi.
    \end{equation}

    After this field redefinition the action becomes:

    \begin{equation}\label{eqn: 2dactionv2}
        S_{2d} = N \int d^2x \ \Big( \frac{u^4 \Lambda^{3/2}}{4} \partial_m X \partial^m X + \frac{1}{8\pi} D_m \varphi D^m \varphi - \frac{1}{8\pi} \epsilon^{mn} \varphi \mathcal{J}_{mn} \Big)
    \end{equation}

    where we have defined a gauge-invariant derivative $D_m \varphi = \partial_m \varphi + a_m$. 
    The field strength $\mathcal{J}_{mn}$ can be equivalently described using the $j_m$ or $a_m$ fields since they differ by a total derivative. 
  
    The above action \eqref{eqn: 2dactionv2} can be simplified by choosing a light-cone gauge $a_-=0$ in order to arrive at the following simple action 
    
    \begin{equation}\label{eqn: 2dactionv3}
        S_{2d} = N \int d^2x \ \Big( \frac{u^4 \Lambda^{3/2}}{4} \partial_m X \partial^m X + \frac{1}{8\pi} \partial _m \varphi \partial ^m \varphi \Big) \, .
    \end{equation}

As we shall see, the above action \eqref{eqn: 2dactionv3} coincides with the composite wall action for $p'-p=1$.

    \subsection{Massive translational modes}\label{sec: translational modes}

    So far we have taken the same centre of mass position $X$ for the scalar and gauge components of the domain wall, because this is required in order to satisfy the Bogomol'nyi equations. This suggests that separating the components comes with an associated energy cost, and we can plot this to find a potential as a function of the difference in centre of mass positions.
    
    The final term in the Bogomol'nyi factorisation (the topological part depending only on the boundary conditions at $\pm \infty$) is independent of the components' centre of mass positions. This means we need to consider the other terms in the Bogomol'nyi factorisation to see how the wall tension changes. The term mixing between the gauge and scalar fields is:
    
    \begin{equation}\label{eqn: BPS mixing term}
        N \left(\rho A'_y - \frac{\partial_x A'_y}{4 \pi \rho} \right)^2. 
    \end{equation}

    Setting this term to zero gives us one of the Bogomol'nyi equations. Substituting in the domain wall solutions necessarily causes this term to vanish, but if we take different centre of mass positions then we will see an increase from zero, because doing so takes us away from the Bogomol'nyi limit. We can see this by substituting in the solutions after taking a thin wall limit:

    \begin{equation}
        \rho = u \  \vartheta(x+X_s) \qq A'_y = j_y \vartheta(-x-X_g)
    \end{equation}
    
    where $\vartheta (x)$ is the Heaviside step function and we have given the scalar and gauge components different centre of mass positions $X_s$ and $X_g$. \eqref{eqn: BPS mixing term} becomes:

    \begin{equation}
        N \left( u  j_y \vartheta(x+X_s) \vartheta(-x-X_g) + \frac{j_y \delta(-x-X_g)}{4 \pi u \vartheta(x+X_s)} \right)^2
    \end{equation}

    which is minimised when $X_s=X_g$: the first term increases linearly when $X_g<X_s$, and the second term is infinite when $X_g>X_s$.
    
\begin{figure}
    	\centering
    	\includegraphics[width = 4in]{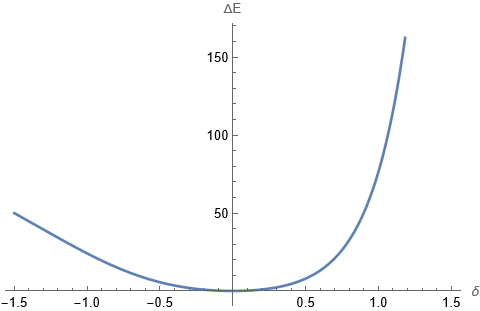}
    	\caption{The energy gain over the tension at the Bogomol'nyi limit, as a function of $\delta = X_s-X_g$, the separation between scalar and gauge components of the wall. Note that the potential is not symmetric, as expected from the thin wall considerations. The plot is for $N=1$.}
   	\label{fig: XY potential}
    \end{figure}

    To find the form of the potential we should consider instead the profiles in \eqref{eqn: fund solns}. We can then perform the integral over $x$ for different separations to find the energy gain as a function of $X_s-X_g$. The integral is complicated, but we can plot the result numerically, as shown in Figure \eqref{fig: XY potential}. Allowing the components' centre of mass positions to fluctuate independently, we might expect to find a massive mode localised on the wall corresponding to the $X_s-X_g$ separation, the mass of which can be obtained by Taylor expanding the potential around its minimum and reading off the coefficient of the quadratic piece. If we set $u=1$, and express the gauge field solution in terms of the electric charge $Q$, then doing this gives us the following potential:

    \begin{equation}
        V(\delta) = \frac{1}{N} \left( \frac{4\pi^2 Q^2 \Lambda^3 (4 \pi^2 + \pi \Lambda^{3/2} + 3 \Lambda^3 )}{(\pi + \Lambda^{3/2}) (2 \pi + \Lambda^{3/2}) (4\pi + 3 \Lambda^{3/2})} \Big( \frac{\delta}{N} \Big)^2 + \frac{8 \pi^2 Q^2 \Lambda^6}{(2\pi + \Lambda^{3/2})(4\pi + 3 \Lambda^{3/2})} \Big( \frac{\delta}{N} \Big)^3 + \mathcal{O}(\delta^4) \right)
    \end{equation}

    where $\delta = X_s - X_g$ and the $\bar{X} = (X_s + X_g)/2$ mode remains massless. The higher-order terms are increasingly suppressed and so can be neglected, however in the $N \rightarrow \infty$ limit the potential vanishes in its entirety and the scalar and gauge components of the wall are free to move independently.

\section{Composite wall, $k\ge N_f/2$}\label{sec: composite wall}

    If we want to consider more complicated domain walls that can interpolate between two arbitrary vacua - those labelled by $p$ and $p'$ for example, where $p$ counts the number of (bosonic meson) eigenvalues in the symmetric minimum - then we need to introduce more scalar fields, and increase the size of our gauge group in order to allow for the maximum amount of Higgsing. Solitons in non-abelian Chern-Simons Higgs models have also been analysed in eg. \cite{Cugliandolo:1989he, Cugliandolo:1990nb, Lee:1990ep}.
    
    We upgrade our action to

    \begin{equation}\label{eqn: 3d action}
        S = N \int d^3x \ \mathrm{tr} \Big( D^\mu \Phi (D_\mu \Phi)^\dagger + \frac{1}{4 \pi} \epsilon^{\mu\nu\rho}( A_\mu \partial_\nu A_\rho + \frac{2i}{3} \  A_\mu A_\nu A_\rho) - \Lambda^3 | \Phi |^2 (|\Phi|^2 - u^2 \mathbb{I} )^2 \Big)
    \end{equation}

    where $A_\mu$ is the $U(\tilde{N})$ gauge field and $\Phi$ is an $N_f \times \tilde{N}$ scalar field transforming in the fundamental of $U(\tilde{N})$ and also transforming in the fundamental of a $U(N_f)$ flavour symmetry. The trace is taken over gauge indices, but the first and last terms require an additional trace over flavour indices to be performed. All traces in this section, where unspecified, should be performed in this way. The prefactor of the cubic part of the Chern-Simons term is taken so that in the pure gauge case we recover the $\epsilon^{\mu\nu\rho} F_{\nu\rho} = 0$ equation of motion, where the field strength is defined as $F_{\mu\nu} = \partial_\mu A_\nu - \partial_\nu A_\mu +i [A_\mu, A_\nu]$. In this section we take $\tilde{N} \geq N_f$ which corresponds to $k \geq N_f/2$ in the parameters of the original (fermionic) theory.

    Using QR matrix decomposition we can rewrite our scalar field like $\Phi = g \rho$ where $g$ is an element of $U(\tilde{N})$ and $\rho$ is an upper triangular $N_f \times \tilde{N}$ matrix with real diagonal entries. We can count degrees of freedom. $\rho$ has $N_f (N_f -1)/2$ complex degrees of freedom and $N_f$ real degrees of freedom. $g$ has $\tilde{N}^2$ degrees of freedom, but not all of them are realised in $\Phi$ because some will only multiply the zero entries in $\rho$. The bottom right $\tilde{N}-N_f$ square of $g$ contains these non-contributing elements so we are left with $\tilde{N}^2 - (\tilde{N}-N_f)^2$ physical degrees of freedom. Between $\rho$ and $g$ this gives us $2 N_f \tilde{N}$ physical degrees of freedom, which is what we expect from a complex $N_f \times \tilde{N}$ scalar field.

    The fields transform under a gauge transformation like:

    \begin{equation}
        \Phi \rightarrow U \Phi \qq \qq A_\mu \rightarrow U A_\mu U^\dagger - i U \partial_\mu U^\dagger
    \end{equation}

    where $U \in U(\tilde{N})$ so that the covariant derivative $D_\mu\Phi = \partial_\mu \Phi + i A_\mu \Phi$ transforms in the correct way. Splitting up $\Phi$ into its constituents we can rewrite this like:

    \begin{align}\label{eqn: GI field definition}
        D_\mu \Phi & = \partial_\mu g \rho + g \partial_\mu \rho + i A_\mu g \rho \nonumber \\ & = g ( g^\dagger \partial_\mu g \rho + \partial_\mu \rho + i g^\dagger A_\mu g \rho) \nonumber \\ & = g (\partial_\mu \rho + i A'_\mu \rho)
    \end{align}

    where we have defined $A'_\mu = g^\dagger A_\mu g - i g^\dagger \partial_\mu g$. If we take the gauge transformation to act on the scalar field constituents like $g \rightarrow U g$ and $\rho \rightarrow \rho$ then $A'_\mu$ is gauge invariant. This is analogous to the method we used in the Abelian case.

    We can then rewrite the action as:

    \begin{equation}\label{eqn: NA redefined action}
        S = N \int d^3 x \ \mathrm{tr} \Big( D'_\mu \rho (D'^\mu \rho)^\dagger + \frac{1}{4\pi} \epsilon^{\mu\nu\rho}( A'_\mu \partial_\nu A'_\rho + \frac{2i}{3} A'_\mu A'_\nu A'_\rho) - \Lambda^3 \rho \rho^\dagger (\rho \rho^\dagger - u^2 \mathbb{I})^2 \Big)
    \end{equation}

    plus the boundary terms:

    \begin{equation}\label{eqn: NA boundary terms}
        S_b = N \int d^3 x \ \frac{1}{4\pi} \epsilon^{\mu\nu\rho} \ \mathrm{tr}_\mathrm{gauge} \Big( - i \partial_\nu ( g^\dagger \partial_\mu g A'_\rho ) + \frac{1}{3} g^\dagger \partial_\mu g g^\dagger \partial_\nu g g^\dagger \partial_\rho g \Big).
    \end{equation}

    The second term is the winding number density, responsible for the quantisation condition on the Chern-Simons level.
    
    Applying the domain wall ansatz $\partial_t = \partial_y = 0$ and making the choice $A'_x = 0$ reduces the action to:
    
    \begin{equation}
        S = N \int d^3 x \ \mathrm{tr} \Big( \partial_x \rho \partial^x \rho^\dagger + A_t A^t \rho \rho^\dagger + A_y A^y \rho \rho^\dagger + \frac{1}{2\pi} A_t \partial_x A_y - \Lambda^3 \rho \rho^\dagger (\rho \rho^\dagger - u^2 \mathbb{I})^2 \Big).
    \end{equation}

    The bosonic meson $\tilde{M} = \Phi^\dagger \Phi = \rho^\dagger \rho$ is gauge invariant but transforms under flavour like $\tilde{M} \rightarrow V^\dagger \tilde{M} V$. This means that we can diagonalise the meson VEV in the vacuum. A general domain wall will interpolate between vacua labelled by $p$ and $p'$, where $p$ and $p'$ count the number of meson eigenvalues in the symmetric (zero) minimum. The meson VEV in the $p$ vacuum is then a diagonal $N_f \times N_f$ matrix with $N_f - p$ entries $u^2$, and $p$ entries $0$. We take $p' > p$.

    Moving between these vacua means that $p'-p$ of the diagonal entries must interpolate between $0$ and $u^2$. All other entries must assume the same values in the vacua at $\pm \infty$. In principle any or all of the meson matrix components can switch on (or condense) on the domain wall, but we will find that only the $p'-p$ interpolating diagonal entries have non-trivial profiles.

    The energy of the system is:

    \begin{equation}\label{eqn: NA energy}
        E = N \int d^2 \boldsymbol{x} \ \mathrm{tr} \Big( \partial_x \rho^\dagger \partial_x \rho + \Lambda^3 \rho^\dagger \rho (\rho^\dagger \rho - u^2 \mathbb{I} )^2 + \rho \rho^\dagger A'_t A'_t + \rho \rho^\dagger A'_y A'_y \Big).
    \end{equation}

    Now we want to find a Bogomol'nyi factorisation. We can do this in a similar way to the Abelian case. We will need to employ the $A'_\mu$ equation of motion, which in the non-abelian case is:

    \begin{equation}
        \frac{1}{4\pi} \epsilon^{\mu\nu\rho}F'_{\nu\rho} + 2  \rho \rho^\dagger A'^\mu + i (\rho \partial^\mu \rho^\dagger - \partial^\mu \rho \rho^\dagger) = 0
    \end{equation}

    with the $t$ component reducing under our domain wall ansatz to:

    \begin{equation}\label{eqn: NA At EoM}
        \frac{1}{2\pi} \partial_x A'_y + 2  \rho \rho^\dagger A'^t = 0 \qq \qq \implies \qq \qq \frac{1}{4\pi} \rho^{-1} \partial_x A'_y = - \rho^\dagger A'^t.
    \end{equation}

    Multiplying the above equation by its Hermitian conjugate allows us to rewrite the $\rho \rho^\dagger A'_t A'_t$ term in terms of $A'_y$. Doing this the energy can be written as:

    \begin{align}\label{eqn: NA BPS factorisation}
        E = N\int d^2 \boldsymbol{x} \ \mathrm{tr} \Big( |\partial_x \rho \mp \Lambda^{3/2} \rho (\rho^\dagger \rho - u^2 \mathbb{I} )|^2 + | \frac{1}{4 \pi } \partial_x A_y (\rho^{\dagger})^{-1} \mp A_y \rho|^2  \Big) \nonumber \\  \pm N L_y \ \mathrm{tr} \Big[ \frac{\Lambda^{3/2}}{2} ( \rho \rho^\dagger - u^2 \mathbb{I})^2 + \frac{1}{4 \pi} A_y^2 \Big]^{x=+\infty}_{x=-\infty}
    \end{align}

    where performing the integral over the $y$ direction gives us the volume factor $L_y$. From this we can read off the Bogomol'nyi equations:

    \begin{align}
        & \partial_x \rho - \Lambda^{3/2}\rho ( \rho^\dagger \rho - u^2 \mathbb{I}) = 0 \label{eqn: PhiBPS} \\ & \frac{1}{4 \pi} \partial_x A'_y - A'_y \rho \rho^\dagger = 0 \label{eqn: AyBPS}
    \end{align}

    where we have taken the upper signs. The scalar field Bogomol'nyi equation can be rewritten in terms of the bosonic meson, which is real-valued, like:

    \begin{equation}
        \partial_x \tilde{M} = 2 \Lambda^{3/2} \tilde{M} ( \tilde{M} - u^2\mathbb{I}).
    \end{equation}

    The boundary conditions discussed above equation \eqref{eqn: NA energy} give rise to interpolating solutions for the $p'-p$ interpolating eigenvalues of the form:

    \begin{equation}\label{eqn: meson evalue soln}
        \tilde{M}_i = u^2 (1 + e^{2 u^2 \Lambda^{3/2} (x+X_i)})^{-1}
    \end{equation}

    with the other diagonal entries remaining constant. Note that each of the $p'-p$ interpolating eigenvalues come with their own centre of mass position $X_i$. Now we should switch on the off-diagonal components as a small perturbation and check whether they can condense on the wall or not.

    Keeping only terms up to first order in the perturbations we find the following Bogomol'nyi equations for the off-diagonal elements:

    \begin{equation}
        \partial_x \tilde{M}_{ij} = 2 \Lambda^{3/2} ( \tilde{M}_i + \tilde{M}_j - u^2 ) \tilde{M}_{ij}
    \end{equation}

    with $\tilde{M}_i$ and $\tilde{M}_j$ referring to the diagonal elements in the same row/column as the off-diagonal element $\tilde{M}_{ij}$. Using the solutions \eqref{eqn: meson evalue soln} for the diagonal elements we can solve the off-diagonal equation, but the factor of $-u^2$ in the parentheses gives an exponential contribution and the only solution respecting the boundary conditions is the trivial (zero) solution.

    From this we conclude that the Bogomol'nyi equations are solved only for a diagonal bosonic meson matrix. What does this mean for $\rho$ though? The original scalar field $\Phi$ can be diagonalised in the vacuum using combined gauge and flavour transformations. The decomposition $\Phi = g \rho$ then splits it into two diagonal matrices, with $g$ containing phases and $\rho$ containing magnitudes, which means that $\rho$ is real-valued.

    This means that $g$ is restricted to group elements of $U(1)^{\tilde{N}}$ with only the $U(1)^{N_f}$ subgroup contributing physical phases. This restriction is appropriate for a domain wall solution which is multiple copies of the fundamental wall, and at this point, before considering the second Bogomol'nyi equation, this appears to be what we have.

    Again it may be possible for off-diagonal components of $\rho$ to have non-trivial profiles on the wall. The Bogomol'nyi equation for $\rho$, in components, is:

    \begin{equation}
        \partial_x \rho_{ij} - \Lambda^{3/2}\rho_{ij} ( \tilde{M}_j - u^2 \mathbb{I}) = 0.
    \end{equation}

    The entries of $\rho$ obey the same boundary conditions as those of the bosonic meson. Considering the off-diagonal elements, which must return to zero at $\pm \infty$, there are three possibilities for the Bogomol'nyi equation: the relevant meson component $\tilde{M}_j$ is either $0$, $u^2$, or has an interpolating profile. These three cases correspond to exponential, constant, and interpolating solutions for $\rho_{ij}$ respectively, and the only one of these which satisfies the boundary conditions is the second, with the constant being zero. We conclude that $\rho$ stays diagonal (and therefore real) everywhere. Its diagonal elements have profiles:

    \begin{equation}
        \rho_i = u (1 + e^{2 u^2 \Lambda^{3/2} (x+X_i)})^{-1/2}.
    \end{equation} 

    Now we want to solve the second Bogomol'nyi equation:

    \begin{equation}
        \frac{1}{4 \pi} \partial_x A'_y = A'_y \rho \rho^\dagger.
    \end{equation}

    Since we now have the profile for $\rho$, we know that $\rho \rho^\dagger$ is an $\tilde{N} \times \tilde{N}$ diagonal matrix with its first $N_f$ entries the same as those of the bosonic meson $\tilde{M}$, and its remaining $\tilde{N}-N_f$ entries zero. We can now proceed to solve the gauge field equation.
    
    We start by dividing up $\rho \rho^\dagger$ like:

    \begin{equation}
        \rho \rho^\dagger = \begin{pmatrix} u^2\mathbb{I} & 0 & 0 \\ 0 & d & 0 \\ 0 & 0 & 0\end{pmatrix} \ \begin{matrix} \}  \\ \}  \\ \}  \end{matrix} \ \begin{matrix}  N_f-p' \\  p'-p \\  \tilde{N} - N_f +p \end{matrix}
    \end{equation}

    where $d$ is a $(p'-p)\times(p'-p)$ matrix containing the interpolating $\tilde{M}_i$s along its diagonal.

    Now we divide up $A'_y$ in a similar way:

    \begin{equation}
        A'_y = \begin{pmatrix} A_{1} & A_{2} & A_{3} \\ A_{2}^\dagger & A_{4} & A_{5} \\ A_{3}^\dagger & A_{5}^\dagger & A_{6} \end{pmatrix} \ \begin{matrix} \}  \\ \}  \\ \}  \end{matrix} \ \begin{matrix}  N_f-p' \\  p'-p \\  \tilde{N} - N_f + p \end{matrix}
    \end{equation}

    where the sub-matrices $A_i$ should more properly all come with $y$ indices and primes but we suppress them to avoid cluttering the notation. From here onward all sub-matrices with lowered $ijk$ indices should be though of as those belonging to $A'_y$. The Bogomol'nyi equation becomes:

    \begin{equation}
        \frac{1}{4 \pi} \begin{pmatrix} \partial_x A_{1} & \partial_xA_{2} & \partial_x A_{3} \\ \partial_x A_{2}^\dagger & \partial_x A_{4} & \partial_x A_{5} \\ \partial_x A_{3}^\dagger & \partial_x A_{5}^\dagger & \partial_x A_{6} \end{pmatrix} =  \begin{pmatrix} A_{1} & A_{2} d & 0 \\ A_{2}^\dagger & A_{4} d & 0 \\ A_{3}^\dagger & A_{5}^\dagger d & 0 \end{pmatrix}
    \end{equation}

    from which we can immediately read off:

    \begin{equation}
        \partial_x A_{3} = \partial_x A_{5} = \partial_x A_{6} = 0.
    \end{equation}

    The conjugate equations for $A_{3}$ and $A_{5}$ are:

    \begin{equation}
        \frac{1}{4\pi} \partial_x A_{3}^\dagger =  A_{3}^\dagger = 0 \qq \qq \frac{1}{4\pi} \partial_x A_{5}^\dagger =  A_{5}^\dagger d = 0
    \end{equation}

    where the equality to zero follows from the fact that $A_{3}$ and $A_{5}$ are constant. This means that $A_{3}=A_{5}=0$ everywhere, in order to satisfy the Bogomol'nyi equation. The $A_{2}$ equation and its conjugate are:

    \begin{equation}
        \frac{1}{4\pi} \partial_x A_{2} =  A_{2} d \qq \qq \frac{1}{4\pi} \partial_x A_{2}^\dagger =  A_{2}^\dagger
    \end{equation}

    which means $A_{2}d_{p'-p}=A_{2}$ and because this must hold everywhere, we conclude that $A_{2}=0$. The equation for $A_{1}$ can be solved:

    \begin{equation}
        \frac{1}{4\pi} \partial_x A_{1} =  A_{1} \quad \implies \quad A_{1} = Be^{4\pi x}
    \end{equation}

    with the coefficient $B$ determined by the boundary conditions.

    To progress from here then we need to consider the boundary conditions. From equation \eqref{eqn: NA energy} we know that $\rho \rho^\dagger A'_yA'_y$ must vanish at $\pm \infty$ in order for the configuration to have finite energy. Separating this out into components, and applying the solutions we already have, we find:

    \begin{align}
        A_{1}^2 = 0 \quad\quad & \mathrm{in \ the} \ p \ \mathrm{vacuum}  \\ A_{1}^2=0, \quad A_{4}^2=0 \quad & \mathrm{in \ the} \ p' \ \mathrm{vacuum}
    \end{align}

    where we have used the fact that $d=0$ in the $p$ vacuum. These boundary conditions mean that $A_{1}=0$ everywhere (the coefficient $B=0$), and since $A_{6}$ makes no appearance, we conclude that it can take any (constant) value. This unconstrained $(\tilde{N} - N_f +p) \times (\tilde{N}-N_f +p)$ sub-matrix realises the remaining unbroken $U(k+p-N_f/2)$ gauge group in the $p$ vacuum (once we recall that $\tilde{N}=k+N_f/2$).

    This leaves us with just one sub-matrix equation left to solve:

    \begin{equation}\label{eqn: small_BPS}
        \frac{1}{4\pi} \partial_x A_{4} =  A_{4} d
    \end{equation}

    where both sides are now $(p'-p) \times (p'-p)$ matrices. We can express $A_{4}$ as a linear combination of the generators $T^a$ of $U(p'-p)$, like $A_{4}=A_4^aT^a$ where $a$ runs from $1$ to $(p'-p)^2$. The gauge coefficients $A_4^a$ will be labelled with an upper $abc$ index. The convention for the generators that we'll adopt divides the $(p'-p)^2$ generators into diagonal generators for the $U(1)^{p'-p}$ maximal Abelian subgroup, which we call $T^D$, plus 'off-diagonal' generators which we call $T^R$ and $T^Q$. The forms of these generators are given below. They obey the usual $\mathrm{tr}(T^aT^b)=\delta^{ab}/2$ relation.

    The diagonal generators look like:

    \begin{equation}
        T^{D} = \frac{1}{\sqrt{2}} \begin{pmatrix} 1 & & & & \\ & 0 & & & \\ & & \ddots & \\ & & & 0 \\ & & & & 0  \end{pmatrix} 
    \end{equation}

    which have a single $1$ in one of the diagonal entries. There are therefore $p'-p$ of this type of generator. The off-diagonal generators look like:

    \begin{equation}
        T^R = \frac{1}{2} \begin{pmatrix} 0 & & & & \\ & \ddots & & & 1 \\ & & 0 & & \\ & & & 0 & \\ & 1 & & & 0  \end{pmatrix} \qq \qq T^Q = \frac{1}{2} \begin{pmatrix} 0 & & & & \\ & \ddots & & & -i \\ & & 0 & & \\ & & & 0 & \\ & i & & & 0  \end{pmatrix}
    \end{equation}

    so we have $(p'-p)(p'-p-1)/2$ of each type of generator - one for each component of the upper/lower triangle where we can place a $1$ or $i/-i$ pair.

    For simplicity let's consider the case of $N_f = 2$, with both meson eigenvalues interpolating according to the solution in \eqref{eqn: meson evalue soln}. The interpolating scalar sub-matrix $d$ and the $A_4$ sub-matrix of the $A'_y$ field look like:

    \begin{equation}
        d = \begin{pmatrix} \tilde{M}_1 & 0 \\ 0 & \tilde{M}_2 \end{pmatrix} \qq\qq A_4 = \frac{1}{2} \begin{pmatrix} \sqrt{2} A_4^1 & A_4^3 - i A_4^4 \\ A_4^3 + i A_4^4 & \sqrt{2} A_4^2 \end{pmatrix} 
    \end{equation}

    so there are four gauge coefficients $A_4^a$ that we need to solve for. The Bogomol'nyi equation is:

    \begin{equation}
        \frac{1}{4\pi} \begin{pmatrix} \sqrt{2} \partial_x A_4^1 & \partial_x (A_4^3 - i A_4^4) \\ \partial_x (A_4^3 + i A_4^4) & \sqrt{2} \partial_x A_4^2 \end{pmatrix} =  \begin{pmatrix} \sqrt{2} A_4^1 \tilde{M}_1 & (A_4^3 - i A_4^4) \tilde{M}_2 \\ (A_4^3 + i A_4^4) \tilde{M}_1 & \sqrt{2} A_4^2 \tilde{M}_2 \end{pmatrix}
    \end{equation}

    The diagonal equations can be solved easily. They have the solutions:

    \begin{equation}\label{eqn: diagonal gauge soln}
        A_4^a = j_a (1+e^{-2u^2\Lambda^{3/2}(x + X_a)})^{-2\pi/ \Lambda^{3/2}}
    \end{equation}

    where the centre of mass position $X_a$ is inherited from the corresponding bosonic meson eigenvalue solution. At the moment these positions are allowed to all be different, but this will change when we consider the off-diagonal equations. Note that each gauge coefficient is allowed by the boundary conditions to have its own $j_a$ value independent of the others. After solving for all the diagonal gauge coefficients what we have corresponds to $p'-p$ copies of the fundamental wall - each scalar component comes with a $U(1)$ gauge field.

    Next we'll look at the off-diagonal equations. Adding the pair of equations which are diagonally opposite gives:

    \begin{equation}
        \frac{1}{2\pi} \partial_x A_4^3 =  (\tilde{M}_1 + \tilde{M}_2) A_4^3 + i  (\tilde{M}_1 - \tilde{M}_2) A_4^4
    \end{equation}

    so in order for the imaginary parts to match on both sides we require either $A_4^4 = 0$ or $\tilde{M}_1 = \tilde{M}_2$. If we assume the second option we find the same interpolating solutions as in equation \eqref{eqn: diagonal gauge soln} for the off-diagonal elements. The generalisation to $p'-p > 2$ works in exactly the same way.

    This means that if all of the scalar components' centre of mass positions coincide, the Bogomol'nyi equations admit a solution where the full $U(p'-p)$ gauge field fluctuates on the wall. In contrast, when all of the scalar components have different centre of mass positions, the only solution to the Bogomol'nyi equations is that of $p'-p$ free, non-interacting fundamental walls, each of which carries a $U(1)$ gauge field. If we try to split up a composite wall by giving the scalar components different centre of mass positions then the off-diagonal components of the $U(p'-p)$ gauge field (the W bosons) become massive and will be absent from the low-energy description. The energy scale below which they can be integrated out (the mass of the W bosons) corresponds to the energetic cost in separating the fundamental walls, which can be found by examining in a similar way to Section \eqref{sec: translational modes} the behaviour of the bulk terms in the Bogomol'nyi factorisation \eqref{eqn: NA BPS factorisation} away from the Bogomol'nyi limit.

    In general we can have solutions to the Bogomol'nyi equations where the gauge field takes an arbitrary block diagonal form (within the $p'-p$ square sub-matrix), with the size of the blocks corresponding to the number of coincident scalar components. Each block of size $n \leq p'-p$ corresponds to a wall carrying a $U(n)$ gauge field. The walls are free and non-interacting, as long as we take all other gauge field components zero.

    For now let's consider the (maximally) composite wall. Taking all of the scalar components to have the same centre of mass position means that the wall carries a $U(p'-p)$ gauge field. While the transverse positions $X_i$ of all the wall components are fixed to the same value, the amplitudes $j_a$ (which are integration constants) are not, and can take any value, like the $j$ appearing in the Abelian case, equation \eqref{eqn: fund solns}, except now there are $(p'-p)^2$ of them, one for each generator.

    The next step is to find $\mathrm{tr}(A_y^2)$ in order to evaluate the tension from equation \eqref{eqn: NA BPS factorisation}. We only need to find the trace of $(A_{4})^2$ because the other sub-matrices are either zero or constant. Because $A_{4}$ is Hermitian, $\mathrm{tr}(A_y^2)$ is just the sum of its entries squared, with the generator normalisation applied:

    \begin{equation}
        \mathrm{tr}_\mathrm{gauge} (A_y^2) = \frac{1}{2} \sum_{a=1}^{(p'-p)^2} (A_4^a)^2.
    \end{equation}

    This means in evaluating the tension we just need to sum the squares of the gauge coefficients $j_a$. The wall tension from equation \eqref{eqn: NA BPS factorisation} is then:

    \begin{equation}
        T = N \ \left[\frac{1}{8\pi} \sum_{i=1}^{(p'-p)^2} j_i^2 + \frac{\Lambda^{3/2}}{2}(p' - p) \right].
    \end{equation}
    
    The tension goes as $N$, and has both linear and non-linear dependence on $(p'-p)$. This expression reproduces the Abelian result, equation \eqref{eqn: abelian tension}, once we account for the different generator normalisations.

    We should check also what becomes of $A'_t$, which in the Abelian case we found had the same profile as $A'_y$. This time each is an $\tilde{N} \times \tilde{N}$ matrix, but by using the $A'_\mu$ equation of motion, equation \eqref{eqn: NA At EoM}, and substituting in our expressions for the $A'_y$ and $\tilde{M}$ components, we recover the same result, that $A'_t=A'_y$ (with indices lowered). Importantly, this means that the collection of $j_a$ parameters is the same for each.

    In this section we have found multiple solutions to the Bogomol'nyi equations corresponding to how many of the scalar components are aligned. The maximally composite wall has all scalar components with the same centre of mass position, and a $U(p'-p)$ gauge field on the wall. At the other extreme we can take all scalar components to have different centre of mass positions, which leaves us with $p'-p$ fundamental walls, each of which carries a $U(1)$ gauge field. The tension has a linear contribution from the interpolating scalar components, and a non-linear part from the gauge field which contributes a $j_a^2$ for each gauge coefficient. This means that the collection of fundamental walls is the solution to the Bogomol'nyi equations with the lowest energy.

    The non-abelian magnetic field is given by $B = -F_{xy} = - g F'_{xy} g^\dagger =  - g  \partial_x A'_y g^\dagger$ which is not gauge-invariant. We are therefore interested in $\mathrm{tr}(B) = \mathrm{tr}(- \partial_x A'_y)$. After performing a canonical rescaling $\rho \rightarrow \rho/\sqrt{N}$ the $A'_t$ equation of motion \eqref{eqn: NA At EoM} becomes:

    \begin{equation}
        \frac{N}{2\pi} \partial_x A'_y = -2 \rho \rho^\dagger A'^t \qq \implies \qq - \frac{N}{2\pi} B = J^t
    \end{equation}

    where we have defined the non-abelian electric current $J^t = -2 \rho \rho^\dagger A'^t$. Integrating both sides over $x$ and taking the trace relates the non-abelian magnetic flux and electric charge carried by the domain wall by

    \begin{equation}
        \mathrm{tr}(j)= \frac{2\pi}{N}  \mathrm{tr}(Q)
    \end{equation}
    
    where $j$ and $Q$ refer this time to $\tilde{N}\times \tilde{N}$ matrices. The magnetic flux matrix contains a central $p'-p$ square block containing the $j_a$ parameters from the gauge coefficient profiles in \eqref{eqn: diagonal gauge soln}, and has the rest of its entries zero. Each side of the relation is just the sum of $p'-p$ lots of $U(1)$ fluxes/charges, with each related to its partner by $j_a = 2\pi Q_a / N$. If we fix the entries of $Q$ to be all $\mathcal{O}(1)$ then the $j_a$ parameters will be $1/N$ suppressed.

    We can write the wall tension in terms of the non-abelian electric charge matrix like

    \begin{equation}
        T = N \ \left[\frac{\pi}{2 N^2} \sum_{i=1}^{(p'-p)^2} q_i^2 + \frac{\Lambda^{3/2}}{2}(p' - p) \right].
    \end{equation}

    where the $q_i$ are the entries of the central (nonzero) $p'-p$ square block within $Q$, which we take to be $\mathcal{O}(1)$. Writing the tension in this way shows that the main contribution is from the scalar part of the wall, which is linear in $p'-p$. The non-linear piece, coming from the gauge field, is suppressed by an overall factor of $1/N$, which means that in the $N \rightarrow \infty$ limit the tension of a composite wall is equal to the tension of $p'-p$ fundamental walls.

    This can be seen using the arguments of section \eqref{sec: translational modes}. There we saw that the interaction between scalar and gauge components is suppressed as $N \rightarrow \infty$, and in the case of the composite wall the interactions between scalar components are mediated by interactions with the gauge components. This means that increasing $N$ allows us to separate the composite wall into fundamental walls, and the tension reduces to that of $p'-p$ fundamental walls, as above.

\section{Field theory on the composite wall}\label{sec: composite wall FT}

    Now we would like to repeat the process of section \eqref{sec: field} to derive the field theory on the composite domain wall's $1 + 1$ dimensional world-volume. We consider the maximally composite wall obtained by interpolating $p'-p$ bosonic meson eigenvalues with the same centre of mass position. Based on eg. \cite{Elitzur:1989nr,Berman_2009,Chu:2009ms} we expect the resulting theory to be some kind of WZW model.

    \subsection{The perturbed solutions}

     We start by taking the domain wall solutions found in the previous section, and upgrading the $X$ and $j$ parameters to fields depending on the wall world-volume coordinates:

    \begin{equation}
        \rho_i = u (1+e^{2u^2 \Lambda^{3/2}(x+X_i(t,y))})^{-1/2} \qq A'^a_{4,m} = j^a_m(t,y) (1+e^{-2 u^2 \Lambda^{3/2}(x+X(t,y))})^{-2\pi/\Lambda^{3/2}}
    \end{equation}

    where $m = t, y$ is a $1+1$ dimensional spacetime index and $a \in (1, ... , (p'-p)^2)$ labels the gauge component. We allow the scalar components to fluctuate independently around the overall centre of mass position $\bar{X} = \sum_{i=1}^{p'-p} X_i /(p'-p)$ but since the gauge components won't contribute any $\partial X$ terms to the 2d action we are free to keep them all at the same position $X$. We will find that each scalar centre of mass $X_i$ will come with its own kinetic term in the 2d action.

    In the non-abelian case there are extra fields that we should switch on, on top of the profile parameters. The general rule is to fluctuate everything possible which does not cost an infinite amount of energy, and to fluctuate those parameters that take us away from the Bogomol'nyi limit only by a small amount. In addition we are only interested in fluctuations which localise on the domain wall. For example we could fluctuate the off-diagonal components of $\rho$ around zero by giving them the profile $\rho_{ij}(t,y) \delta(x+X(t,y))$ with $\rho_{ij}$ small. This costs a finite amount of energy, however there is no energetic requirement for the delta function to be located at $x=-X$ over any other position. We should therefore ignore this kind of fluctuation.

    On the other hand, there are off-diagonal components of $A'_4$ which we can turn on by giving them a profile like $k(t,y) \vartheta(-x-X(t,y))$ (with $k$ small) and using the same kind of considerations as in section \eqref{sec: translational modes} we find that the energy of these fluctuations is minimised when they are located at $x=-X$. So these modes are localised at the domain wall and may be important in the 2d theory.

    The form of our perturbed solutions is then (schematically)

    \begin{equation}\label{eqn: NA fluctuations}
        \rho \rho^\dagger = \begin{pmatrix} \mathbb{I} & 0 & 0 \\ 0 & d & 0 \\ 0 & 0 & 0 \end{pmatrix} \ \begin{matrix} \}  \\ \}  \\ \}  \end{matrix} \ \begin{matrix}  N_f-p' \\  p'-p \\  \tilde{N} - N_f + p \end{matrix} \qq\qq A'_m = \begin{pmatrix} 0 & 0 & 0 \\ 0 & j_m & k_m \\ 0 & k^\dagger_m & c_m \end{pmatrix} \ \begin{matrix} \}  \\ \}  \\ \}  \end{matrix} \ \begin{matrix}  N_f-p' \\  p'-p \\  \tilde{N} - N_f + p \end{matrix}
    \end{equation}

    where $j$ is allowed to fluctuate only on one side of the wall, and can be large. $k$ is allowed to fluctuate only on the same side as $j$, and must be small, and $c$ can fluctuate everywhere, and can be large. We will also consider $g$ as a group element of the full $U(\tilde{N})$ group, since it does not contribute to the tension, or appear in the Bogomol'nyi factorisation.

    We took $A'_x = 0$ as part of our domain wall ansatz, and as a result the profile has $A'_x = 0$ everywhere. However in deriving the field theory on the wall we should relax this condition and allow it to make small fluctuations around zero. In the same way to $A'_m$, the only components that we are allowed to turn on are:

    \begin{equation}
        A'_x = \begin{pmatrix} 0 & 0 & 0 \\ 0 & j_x & k_x \\ 0 & k^\dagger_x & c_x \end{pmatrix} \ \begin{matrix} \}  \\ \}  \\ \}  \end{matrix} \ \begin{matrix}  N_f-p' \\  p'-p \\  \tilde{N} - N_f + p \end{matrix}
    \end{equation}

    and this time the $j_x$ field as well as the $k_x$ is a small fluctuation on one side of the wall only. Again $c_x$ is allowed to fluctuate freely on both sides of the wall.

    The non-zero $k+p'-N_f/2$ square block at the bottom right of $A'_\mu$ (that contains the $j_\mu,k_\mu,$ and $c_\mu$ fields) we will denote $B'_\mu$, and the smaller $k+p-N_f/2$ square blocks within these containing the $c_\mu$ fields we will denote $b'_\mu$. The larger $B'_\mu$ fluctuates only on one side of the wall, whereas the smaller $b'_\mu$ is allowed to fluctuate everywhere.

    \subsection{The 2d action}
    
    Now we substitute the above perturbed domain wall solutions back into the 3d action \eqref{eqn: NA redefined action}, remembering to include the boundary terms which will be important here. We add boundary terms according to \eqref{eqn: NA boundary terms} which transform under a gauge transformation where $g \rightarrow Ug$ in such a way as to cancel the gauge variation of the bulk Chern-Simons theories on either side of the wall. These will be $U(k+p'-N_f/2)_N$ and $U(k+p-N_f/2)_N$ Chern-Simons theories. Importantly the Chern-Simons levels have the same sign which means that the anomalies from each side will partially cancel.

    The 3d action reduces to the boundary terms:

    \begin{align}\label{eqn: NA 2d action}
        S_{2d} = \ & N \int d^2x \ \frac{u^4 \Lambda^{3/2}}{4} \sum_{i=1}^{p'-p} \partial_m X_i \partial^m X_i + \frac{1}{8\pi} \mathrm{tr}_\mathrm{gauge} \Big[  B'_m B'^m - b'_m b'^m + B'_x B'^x - b'_x b'^x \Big] \nonumber \\ & + N \int d^3 x \ \frac{1}{4\pi} \epsilon^{\mu\nu\rho} \ \mathrm{tr}_\mathrm{gauge} \ \Pi_{k+p'-N_f/2} \Big( - i \partial_\nu ( g^\dagger \partial_\mu g A'_\rho ) + \frac{1}{3 } g^\dagger \partial_\mu g g^\dagger \partial_\nu g g^\dagger \partial_\rho g \Big) \nonumber \\ & - N \int d^3 x \ \frac{1}{4\pi} \epsilon^{\mu\nu\rho} \ \mathrm{tr}_\mathrm{gauge} \ \Pi_{k+p-N_f/2} \Big( - i \partial_\nu ( g^\dagger \partial_\mu g A'_\rho ) + \frac{1}{3 } g^\dagger \partial_\mu g g^\dagger \partial_\nu g g^\dagger \partial_\rho g \Big)
    \end{align}

    and leaves the following bulk terms remaining:

    \begin{align}\label{eqn: NA 3d action}
        S_{3d} =  & \ N \int_{x>-X} d^3x \ \frac{1}{4\pi} \epsilon^{mxn} \ \mathrm{tr} \Big[ B'_m \partial_x B'_n + B'_x \partial_n  B'_m + B'_n \partial_m B'_x + 2i B'_m B'_x B'_n \Big] \nonumber \\  & + N \int_{x<-X} d^3x \ \frac{1}{4\pi} \epsilon^{mxn} \ \mathrm{tr} \Big[  b'_m \partial_x b'_n + b'_x \partial_n  b'_m + b'_n \partial_m b'_x + 2i b'_m b'_x b'_n \Big].
    \end{align}
    
    The terms on the first line of $S_{2d}$ come from the reduction of the 3d action to 2d. The second and third lines are the anomaly-producing terms required to cancel the gauge non-invariance coming from the Chern-Simons theories on either side of the wall\footnote{In this notation it is not entirely clear that this is the case because our bulk Chern-Simons terms are constructed using the gauge-invariant fields. The anomaly cancellation will become more obvious when we switch back to working with gauge fields shortly, but for now it may be better to think of the anomaly cancelling terms as those required to cancel the gauge-noninvariance of the boundary terms appearing during the original switch from gauge fields to gauge-invariant fields.}. We construct these by projecting onto the relevant Lie algebra the boundary terms written using the $U(\tilde{N})$ gauge field and group element $g$. The projection of the gauge field corresponds to taking the bottom right $k + p' - N_f/2$ and $k+p-N_f/2$ square blocks from our solution in equation \eqref{eqn: NA fluctuations}, which are just the $B'_m$ and $b'_m$ fields.
    
    Since the second and third lines of $S_{2d}$ are Lie algebra valued they will combine into a single set of anomaly-producing terms for the coset $U(k+p'-N_f/2)/U(k+p-N_f/2)$\footnote{We recommend the self-contained paper by Isidro and Ramallo for an excellent introduction to coset models \cite{Isidro:1993er}.}.

    The remaining bulk terms are Chern-Simons terms for $U(k+p-N_f/2)$ and $U(k+p'-N_f/2)$ gauge theories, both with level $N$. This is what we expect for the $p$ and $p'$ low-energy theories in each of the vacua. Performing a parity transformation on one side of the wall means that we can work with both Chern-Simons theories on the same side, but possessing opposite levels. We have also isolated the $x$ index so that we can rewrite the Chern-Simons terms like

    \begin{equation}
        \ N \int_{x>-X} d^3x \ \Big( \frac{1}{4\pi} \epsilon^{mxn} \ \mathrm{tr} \Big[ B'_m \partial_x B'_n - B'_x F'_{mn} \Big] -  \frac{1}{4\pi} \epsilon^{mxn} \ \mathrm{tr} \Big[ b'_m \partial_x b'_n - b'_x f'_{mn} \Big] \Big)
    \end{equation}

    where $F'$ and $f'$ are the field strengths for $B'$ and $b'$ respectively. Now we switch to working with the gauge fields $B_m$ and $b_m$ by reversing the initial definition of $A'_\mu$ in terms of the 3d gauge field $A_\mu$:

    \begin{equation}
        B'_m = H^\dagger B_m H - i H^\dagger \partial_m H \qq \mathrm{and} \qq b'_m = h^\dagger b_m h - i h^\dagger \partial_m h
    \end{equation}

    where $H$ and $h$ are group elements of the same groups as their corresponding gauge fields. It is these fields that we should write the anomaly-producing terms with. Expanding the $B'^2$ and $b'^2$ terms in the action, we get:

    \begin{align}
        S_{2d} = \ & N \int d^2x \ \frac{1}{8\pi} \mathrm{tr} \Big[  (B_mB^m - b_m b^m ) - 2i (B_m \partial^m H H^\dagger - b_m \partial^m h h^\dagger) -(H^\dagger \partial_m H H^\dagger \partial^m H - h^\dagger \partial_m h h^\dagger \partial h) \Big] \nonumber \\ & - N \int d^2 x \ \frac{i}{4\pi } \epsilon^{mn} \mathrm{tr} \Big[ \partial_m H H^\dagger B_n - \partial_m h h^\dagger b_n \Big]  \nonumber \\  & + N \int d^3 x \ \frac{1}{12\pi} \epsilon^{\mu\nu\rho} \ \mathrm{tr}  \Big[ (H^\dagger \partial H)^3 - (h^\dagger \partial h)^3 \Big]
    \end{align}

    where $\epsilon^{mn} = \epsilon^{mxn}$ so that $\epsilon^{ty}=+1$ and $\epsilon^{yt}=-1$. We are also ignoring terms involving the fields $X$, $B_x$, and $b_x$ for now to keep expressions shorter, but will replace them later. The 3d action becomes:

    \begin{equation}
        S_{3d} =  \ N \int_{x>-X} d^3x \ \frac{1}{4\pi} \epsilon^{mxn} \ \mathrm{tr} \Big[ B_m \partial_x B_n - B_x F_{mn} \Big]  -   \frac{1}{4\pi} \epsilon^{mxn} \ \mathrm{tr} \Big[ b_m \partial_x b_n - b_x f_{mn} \Big].
    \end{equation}

    We identify two WZW actions for the $H$ and $h$ fields, which means that we can write the 2d part of the action as:

    \begin{align}\label{eqn: WZWs plus terms}
        S_{2d} = & \ S_{WZW}[H]_N - S_{WZW}[h]_N    \nonumber \\  & + N \int d^2x \ \frac{1}{8\pi} \mathrm{tr} \Big[ (B_mB^m - b_m b^m   ) - 2i (B_m \partial^m H H^\dagger - b_m \partial^m h h^\dagger)  \Big] \nonumber \\ & - N \int d^2 x \ \frac{i}{4\pi} \epsilon^{mn} \mathrm{tr} \Big[ \partial_m H H^\dagger B_n - \partial_m h h^\dagger b_n \Big]
    \end{align}

    where the 2d WZW action at level $N$ is defined as

    \begin{equation}
        S_{WZW}[g]_N = -\frac{N}{8\pi} \int d^2 x \ \mathrm{tr} \Big[ g^\dagger \partial_m g g^\dagger \partial^m g \Big] + \frac{N}{12 \pi} \int d^3 x \ \epsilon^{\mu\nu\rho} \  \mathrm{tr} \Big[ g^\dagger \partial_\mu g g^\dagger \partial_\nu g g^\dagger \partial_\rho g \Big].
    \end{equation}

    The combined bulk and boundary actions can be rewritten as a gauged WZW model (see Appendix \eqref{sec: appendix}) for $U(k+p'-N_f/2)_N/U(k+p-N_f/2)_N$. This is exactly the gauged WZW that is advocated for in \cite{Armoni:2015jsa}.

    The $B_x$ and $b_x$ fields are constrained on the boundary by the equations of motion $B_x - \partial_x H H^\dagger = 0$ and $b_x - i \partial_x h h^\dagger = 0$ which means that our 2d action is, after reintroducing the centre of mass modes:

    \begin{equation}
        S_{2d} = N \int d^2 x \ \frac{u^4 \Lambda^{3/2}}{4} \sum_{i=1}^{p'-p} \partial_m X_i \partial^m X_i + S_{gWZW}[H,A]_N
    \end{equation}

    so we have a gauged WZW model, with $H \in U(k+p'-N_f/2), A\in \mathfrak{u}(k+p-N_f/2)$, and a set of decoupled kinetic terms for the scalar components' centre of mass positions. Beyond leading order in $N$ we will find mass terms and further interactions for the $X_i$, with mass $M_i \sim 1/N$, however a single overall centre of mass mode $\bar{X} = \sum_{i=1}^{p'-p} X_i/(p'-p)$ will remain massless.

\section{Domain wall for the case $k<N_f/2$} \label{sec:two_bosonic}

    In section \eqref{sec: composite wall} we found domain wall solutions in a non-abelian $U(\tilde{N})$ Chern-Simons theory with $N_f$ bosonic flavours, and we took $\tilde{N} \geq N_f$, which corresponds to $k \geq N_f/2$ in the parameters of the original fermionic theory. This meant that allowing all of the scalar fields to condense in the symmetry-breaking minimum corresponded to fully Higgsing the gauge group.

    If we want to consider a situation where $k < N_f/2$, then this means that $\tilde{N} < N_f$ and we can fully Higgs the gauge group without needing to condense all of the scalars. From this point we can do no further Higgsing, and this means that we are unable to realise all $N_f+1$ low-energy theories which are supposed to be level-rank dual to the $N_f+1$ low-energy sectors of QCD\textsubscript{3}.
    
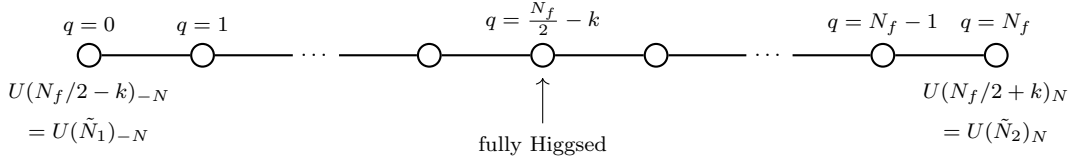
\begin{figure}[t]
        \centering
        
       \begin{scriptsize}
                 \begin{tikzpicture}
                    
                        \node (1) at (-6,1.1) [circle,draw,thick,minimum size=0.3cm,label=above:{$q=0$}] {};
                        \node (13) at (-6,0.6) [] {$U(N_f/2-k)_{-N}$};
                        \node (11) at (-6,0.1) [] {$=U(\tilde{N}_1)_{-N}$};
                        \node (2) at (-4.5,1.1) [circle,draw,thick,minimum size=0.3cm,label=above:{$q=1$}] {};
                        
                        \node (3) at (-3,1.1)  {$\dots$};
                        
                        \node (5) at (-1.5,1.1) [circle,draw,thick,minimum size=0.3cm] {};
                        \node (6) at (0,1.1) [circle,draw,thick,minimum size=0.3cm,label=above:{$q=\frac{N_f}{2}-k$}] {};
                        \node (7) at (1.5,1.1) [circle,draw,thick,minimum size=0.3cm] {};
                        
                        \node (8) at (3,1.1)  {$\dots$};

                        \node (9) at (4.5,1.1) [circle,draw,thick,minimum size=0.3cm,label=above:{$q=N_f-1$}] {};
                        \node (10) at (6,1.1) [circle,draw,thick,minimum size=0.3cm,label=above:{$q=N_f$}] {};
                        \node (14) at (6,0.6) [] {$U(N_f/2+k)_N$};
                        \node (12) at (6,0.1) [] {$=U(\tilde{N}_2)_N$};
                        
                        \draw[thick] (1) -- (2) -- (3) -- (5) -- (6) -- (7) -- (8) -- (9) -- (10);

                        \draw[->]        (0,0.2)   -- (0,0.8);
                        \node (15) at (0,-0.1) [] {fully Higgsed};
                        
                 \end{tikzpicture}
             \end{scriptsize}
      
        \caption{The phase diagram in terms of the two bosonic duals, which are located at either end of the line.}
        \label{fig: phase diagram}
    \end{figure}

    In the $k<N_f/2$ regime we instead need to employ two bosonic duals to cover the entire phase diagram \cite{armoni_2020_metastable}. They are:

    \begin{equation}
        U(N_f/2 - k)_{-N} \qq \mathrm{and} \qq U(N_f/2 + k)_N
    \end{equation}

    with each coupled to $N_f$ flavours of fundamental scalar field, in order to reproduce the desired Grassmannian sigma models in each vacuum. The two theories are located at opposite ends of the 'phase line' (see figure \eqref{fig: phase diagram}) with the other low-energy theories obtained via Higgsing. Fully Higgsing one dual theory leads to the same vacuum as fully Higgsing the other does.

    There are $N_f+1$ vacua, which we will label with $q = (0, 1, \ ... \ , N_f)$. The labelling scheme will be different to usual, where we use $p$ to express the amount of Higgsing done in one 'direction'. This time the 'direction' of our Higgsing needs to switch at the midpoint of the phase line.

    We will take $U(N_f/2 - k)$ to be the $q=0$ vacuum. Then from $q=0$ up to and including $q=N_f/2-k$, the vacuum theories are $U(N_f/2-k-q)$ so that at $q=N_f/2-k$ (which we refer to as the midpoint) we achieve full Higgsing of the gauge group. We want to have the $U(N_f/2+k)$ theory in the $q=N_f$ vacuum, so from $q=N_f/2-k$ up to $q=N_f$ the vacuum theories are $U(N_f/2+k +q-N_f) = U(-(N_f/2-k-q))$. This means that we fully Higgs the second dual as well for $q=N_f/2-k$, as required.

    We will refer to the two gauge groups as $U(\tilde{N}_1)$ and $U(\tilde{N}_2)$ as in figure \eqref{fig: phase diagram}. In order to find domain wall solutions, we will consider a single set of $N_f$ scalar fields transforming in the fundamental representation of $U(\tilde{N}_1)\times U(\tilde{N}_2)$, and then we will break the gauge group down to the desired subgroup (via a suitable Higgsing) in each vacuum.

    A domain wall interpolating between vacua labelled by $q$ and $q'$ with $q, q' \leq N_f/2 - k$ can be obtained in exactly the same way as those found in section \eqref{sec: composite wall}. The same is true for $q, q' \geq N_f/2 -k$. The domain wall solutions novel to this section then are those which interpolate between theories on opposite sides of the phase line midpoint, or in other words those between vacua labelled by $q < N_f/2 -k$ and $q' > N_f/2 -k$. It is this kind of wall therefore that we shall focus on.

    Our action is:

    \begin{align}
        S = N\int d^3x \ \mathrm{tr}\Big[ -\frac{1}{4\pi} \epsilon^{\mu\nu\rho}(A_{1\mu} \partial_\nu A_{1\rho} + \frac{2i}{3} A_{1\mu} A_{1\nu} A_{1\rho} ) + \frac{1}{4\pi} \epsilon^{\mu\nu\rho}(A_{2\mu} \partial_\nu A_{2\rho} + \frac{2i}{3} A_{2\mu} A_{2\nu} A_{2\rho} ) \nonumber \\ + D_\mu \Phi (D^\mu \Phi)^\dagger - \Lambda^3 \Phi^\dagger \Phi (\Phi^\dagger \Phi - u^2)^2 \Big]
    \end{align}

    where $A_1$ is the gauge field for $U(\tilde{N}_1)$ and $A_2$ is the gauge field for $U(\tilde{N}_2)$. We can embed these in a larger $U(\tilde{N}_1 + \tilde{N}_2)$ group like:
    
    \begin{equation}\label{eqn: gauge embedding}
        A = \begin{pmatrix} A_1 & 0 \\ 0 & A_2 \end{pmatrix} \qq \begin{matrix}   \} \\ \}  \end{matrix} \ \begin{matrix} \tilde{N}_1 \\ \tilde{N}_2 \end{matrix}
    \end{equation}

    The scalar field $\Phi$ is charged under the gauge group, and also transforms in the fundamental representation of a $U(N_f)$ flavour group. This means that we can represent it as an $N_f \times (\tilde{N}_1 + \tilde{N}_2)$ matrix, with some off-diagonal components set to zero in the same way as $A$ in the above. We can diagonalise $\Phi$ in the vacuum using combined gauge and flavour transformations, and then in order to achieve the desired Higgsing of the gauge group, we require it to take the following form in the vacua on either side of the wall:

    \begin{equation}
        \Phi_q  = u\begin{pmatrix} 1 & 0 & 0 & 0 \\ 0 & 0 & 0 & 0 \\ 0 & 0 & 1 & 0 \\ 0 & 0 & 0 & 1 \end{pmatrix}  \qq \begin{matrix} \}  \\ \} \\ \} \\ \}  \end{matrix} \ \begin{matrix}  q \\  \tilde{N}_1 -q \\ \tilde{N}_2 - N_f + q' \\ N_f - q' \end{matrix} \qq \qq \Phi_{q'}  = u\begin{pmatrix} 1 & 0 & 0 & 0 \\ 0 & 1 & 0 & 0 \\ 0 & 0 & 0 & 0 \\ 0 & 0 & 0 & 1 \end{pmatrix}  \qq \begin{matrix} \}  \\ \} \\ \} \\ \}  \end{matrix} \ \begin{matrix}  q \\  \tilde{N}_1 -q \\ \tilde{N}_2 - N_f + q' \\ N_f - q' \end{matrix}
    \end{equation}

    We therefore expect a domain wall solution with $q'-q$ interpolating eigenvalues of $\Phi$. We will have $\tilde{N}_1 - q$ eigenvalues interpolating in one direction, and $\tilde{N}_2 - N_f + q'$ interpolating in the opposite direction.

    We perform the usual decomposition of $\Phi$ like $\Phi = g \rho$, where this time $g \in U(\tilde{N}_1 + \tilde{N}_2)$, and we define a gauge-invariant field $A'_\mu$ in the same way as usual (see equation \eqref{eqn: GI field definition}). The procedure to obtain the domain wall solutions then proceeds in exactly the same way as in previous sections.

    The above action has an energy admitting the same Bogomol'nyi completion as in section \eqref{sec: composite wall}, so it has the same Bogomol'nyi equations and the same set of solutions. The scalar field has $q'-q$ interpolating eigenvalues of the form:

    \begin{equation}
        \rho_i = u(1+e^{ \pm 2u^2 \Lambda^{3/2}(x+X)})^{-1/2}
    \end{equation}

    with the $\pm$ corresponding to the direction that the component interpolates in. The gauge field has two interpolating subgroups, for $U(\tilde{N}_1 -q)$ and $U(\tilde{N}_2 - N_f + q')$, with the gauge coefficients taking the usual form:

    \begin{equation}
        A'^a_y = j^a (1+e^{\mp 2u^2 \Lambda^{3/2}(x+X)})^{-2\pi / \Lambda^{3/2}}
    \end{equation}

    where we choose the $\mp$ sign so that the gauge field interpolates in the opposite direction to its corresponding scalar component. All other gauge field entries are zero. This solution to the Bogomol'nyi equations therefore consists of two composite walls interpolating in opposite directions, with nothing binding them together. We might expect some massive W-bosons as remnants of the larger $U(\tilde{N}_1 + \tilde{N}_2)$ group to mediate the interaction between these walls, and indeed we will find these when we consider the effect of fluctuating the off-diagonal gauge components away from the Bogomol'nyi limit.

    Although we have recovered the usual solutions to the Bogomol'nyi equations, the domain wall tension will be different due to some components interpolating oppositely to each other. The tension is given by:

    \begin{equation}
        T = N \ \mathrm{tr} \Big[ \frac{\Lambda^{3/2}}{2} (\Phi^\dagger \Phi - u^2)^2 + \frac{1}{4 \pi} A_{1y}^2 + \frac{1}{4\pi} A_{2y}^2 \Big]^{x=+\infty}_{x=-\infty}
    \end{equation}

    and we will get cancellation between some of the scalar components. The gauge components will will not necessarily lead to cancellation as there is no requirement for the $j$ parameters of the two oppositely interpolating walls to match. Applying the solutions that we have found above, the tension becomes:

    \begin{equation}
        T = N \Big[ \frac{\Lambda^{3/2}}{2} (2k - N_f + q' + q) + \frac{1}{8\pi} ( \sum^{(\tilde{N}_1 - q)^2}_{a=1} j_{2a}^2 - \sum^{(\tilde{N}_2 - N_f + q')^2}_{a=1} j_{1a}^2 ) \Big]
    \end{equation}

    so we have a sum over squares of the $j$ parameters, and a linear piece which however does not depend on $q'-q$ like we might expect it to. Instead it depends on the 'distances' from the midpoint $N_f/2 - k - q$ and $q'-(N_f/2 - k)$.

    \subsection{Field theory on the wall}

    In the same way as in section \eqref{sec: composite wall FT} we will derive the field theory living on the wall's 1+1 dimensional world volume. There will be terms coming from the reduction of the 3d action to 2d, and also anomaly-cancelling terms which this time will need to cancel $U(\tilde{N}_1-q)_{-N}$ and $U(\tilde{N}_2-N_f + q')_N$ anomalies.

    At this point the fact that we are embedding our $U(\tilde{N}_1) \times U(\tilde{N}_2)$ gauge group inside a larger $U(\tilde{N}_1 + \tilde{N}_2)$ group becomes important, because it means that we can consider exciting the zero components in equation \eqref{eqn: gauge embedding}. Doing this will take us away from the Bogomol'nyi limit but as in section \eqref{sec: composite wall FT} if we take small fluctuations localised on the domain wall, then the energy cost is finite and we should consider these fluctuations as fields belonging to the wall's 1+1 dimensional world volume theory.

    These fields are W bosons for $U(\tilde{N}_1 + \tilde{N}_2 )$ and as mentioned earlier we expect them to provide an attractive force between the two components of the wall we constructed above. We can check this numerically using a similar method to section \eqref{sec: translational modes}. Fixing the fluctuations at the two components' centre of mass, as we move the wall components past each other we find a minimum in the energy gain when they coincide. Alternatively we can fix the two components' positions and vary the position of the fluctuations. We find minimal energy gain when the fluctuation shares its position with one of the components, and an overall minimum when all components and fluctuations coincide.

    After fluctuating all fields costing a finite amount of energy, and only considering those which localise on the wall, our solutions then take the schematic form:

    \begin{equation}
        \Phi^\dagger \Phi = \begin{pmatrix} 1 & 0 & 0 & 0 \\ 0 & d(x) & 0 & 0 \\ 0 & 0 & d(-x) & 0 \\ 0 & 0 & 0 & 1 \end{pmatrix} \qq A'_y = \begin{pmatrix} 0 & 0 & 0 & 0 \\ 0 & j(x) & \delta & 0 \\ 0 & \delta & j(-x) & 0 \\ 0 & 0 & 0 & 0  \end{pmatrix}  \qq \begin{matrix} \}  \\ \} \\ \} \\ \}  \end{matrix} \ \begin{matrix}  q \\  \tilde{N}_1 -q \\ \tilde{N}_2 - N_f + q' \\ N_f - q' \end{matrix}
    \end{equation}

    where $A'_y$ is the gauge-invariant field belonging to the Lie algebra of $U(\tilde{N}_1 + \tilde{N}_2)$. We are now in a position to determine the field theory on the wall. We should keep in mind that the $\delta$ fluctuations are small, since they take us away from the Bogomol'nyi limit.
    
    In a similar way to Section \eqref{sec: composite wall FT} we will relabel the non-zero $q'-q$ square sub-matrix of $A'_y$ as $B'_y$, which is a gauge-invariant field valued in the Lie algebra of $U(q'-q)$. The reduction of the 3d action results in the following boundary terms:

    \begin{align}
        S_{2d} = \ & N \int d^2x \ \frac{u^4 \Lambda^{3/2}}{4} \sum_{i=1}^{q'-q} \partial_m X_i \partial^m X_i + \frac{1}{8\pi} \mathrm{tr}_\mathrm{gauge} \Big[  B'_m B'^m  + B'_x B'^x \Big] \nonumber \\ & + N \int d^3 x \ \frac{1}{4\pi} \epsilon^{\mu\nu\rho} \ \mathrm{tr}_\mathrm{gauge} \ \Pi_{k+q'-N_f/2} \Big( - i \partial_\nu ( g^\dagger \partial_\mu g B'_\rho ) + \frac{1}{3 } g^\dagger \partial_\mu g g^\dagger \partial_\nu g g^\dagger \partial_\rho g \Big) \nonumber \\ & + N \int d^3 x \ \frac{1}{4\pi} \epsilon^{\mu\nu\rho} \ \mathrm{tr}_\mathrm{gauge} \ \Pi_{N_f/2 - q - k} \Big( - i \partial_\nu ( g^\dagger \partial_\mu g B'_\rho ) + \frac{1}{3 } g^\dagger \partial_\mu g g^\dagger \partial_\nu g g^\dagger \partial_\rho g \Big).
    \end{align}

    We will denote the projection of $B'$ onto the Lie algebra of $U(N_f/2 - q-k)$ by $b'_1$, and the projection onto the Lie algebra of $U(k+q'-N_f/2)$ by $b'_2$. The associated group elements (which only appear in Lie algebra valued $g^\dagger \partial g$ combinations) we will call $h_1$ and $h_2$. After reducing all possible terms in the action to 2d, the following bulk terms remain:

    \begin{align}
        S_{3d} =  & \ -N \int_{x>-X} d^3x \ \frac{1}{4\pi} \epsilon^{mxn} \ \mathrm{tr} \Big[ b'_{1m} \partial_x b'_{1n} + b'_{1x} \partial_n  b'_{1m} + b'_{1n} \partial_m b'_{1x} + 2i b'_{1m} b'_{1x} b'_{1n} \Big] \nonumber \\  & + N \int_{x<-X} d^3x \ \frac{1}{4\pi} \epsilon^{mxn} \ \mathrm{tr} \Big[  b'_{2m} \partial_x b'_{2n} + b'_{2x} \partial_n  b'_{2m} + b'_{2n} \partial_m b'_{2x} + 2i b'_{2m} b'_{2x} b'_{2n} \Big].
    \end{align}

    The action now takes a form reminiscent of equations \eqref{eqn: NA 2d action} and \eqref{eqn: NA 3d action}. We expect then to be able to rewrite this as some combination of WZW actions. In order to do this we are going to switch back to working with genuine gauge fields, however we will only switch $b'_1$ and $b'_2$, which means that there will be some gauge-invariant terms in the 2d action from the $B'^2$ term which remain. They give mass terms for the (still gauge invariant) W bosons of $U(\tilde{N}_1 + \tilde{N}_2)$.

    Doing this allows us to form two WZW actions, for $U(N_f/2 - q - k)_N$ and $U(k+q'-N_f/2)_N$, with the group elements being $U_1 = u_1 h_1$ and $U_2 = u_2 h_2$ where we have adopted the pure gauge solutions $b_1 = - iu_1^\dagger \partial u_1$ and $b_2 = -i u_2^\dagger \partial u_2$ in the same way as described in Appendix \eqref{sec: appendix}. This means that the field theory on the wall consists of $q'-q$ massless translational modes, along with a $U(N_f/2 - q-k)_{N} \times U(k+q'-N_f/2)_N$ WZW theory, and some massive gauge-invariant fields:
    
    \begin{equation}
        S_{2d} = S_{WZW}
        [U_1]_N + S_{WZW}[U_2]_N + N \int d^2 x \ \left ( \frac{u^4 \Lambda^{3/2}}{4} \sum_{i=1}^{q'-q} \partial_m X_i \partial^m X_i + \frac{1}{4 \pi} \mathrm{tr}\Big[ W'_m W'^{\dagger m} + W'_x W'^{\dagger x} \Big] \right )
    \end{equation}

    and the bulk Chern-Simons terms have become the cubic terms of the WZW actions. We anticipate that as in the case of the composite wall for $k \ge N_f/2$, the fields $X_i$ acquire a mass $M_i \sim 1/N$ while the overall centre of mass mode remains massless.

\section{The domain wall as a D-brane} 
\label{sec: D-brane}

    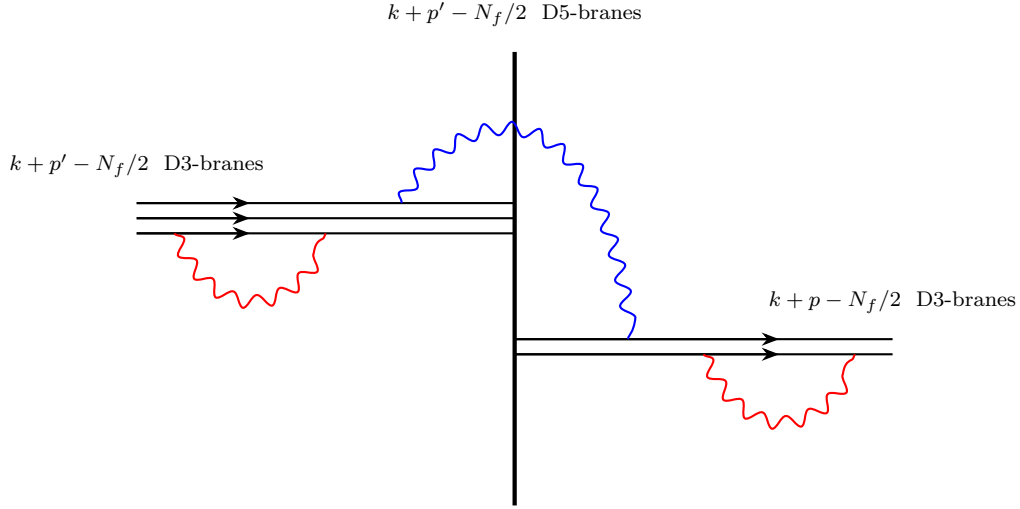
\begin{figure}[t]
        \centering
        
       \begin{scriptsize}
                 \begin{tikzpicture}

                    \node (3) at (0,3.5) [] {$k+p'-N_f/2 \ $ D5-branes};
                    \draw[ultra thick] (0,3) -- (0,-3);

                    \node (1) at (-5,1.5) [] {$k+p'-N_f/2 \ $ D3-branes};
                    \draw[thick] (-5,1) -- (0,1);
                    \draw[-Stealth, thick] (-5,1) -- (-3.5,1);
                    \draw[thick] (-5,0.8) -- (0,0.8);
                    \draw[-Stealth, thick] (-5,0.8) -- (-3.5,0.8);
                    \draw[thick] (-5,0.6) -- (0,0.6);
                    \draw[-Stealth, thick] (-5,0.6) -- (-3.5,0.6);

                    \node (2) at (5,-0.3) [] {$k+p-N_f/2 \ $ D3-branes};
                    \draw[thick] (0,-0.8) -- (5,-0.8);
                    \draw[-Stealth, thick] (0,-0.8) -- (3.5,-0.8);
                    \draw[thick] (0,-1) -- (5,-1);
                    \draw[-Stealth, thick] (0,-1) -- (3.5,-1);

                    \draw[blue, thick, snake it] (-1.5,1) .. controls (-1,2) and (1,3) .. (1.5,-0.8);
                    \draw[red, thick, snake it] (-4.5,0.6) .. controls (-4,-0.5) and (-3,-0.5) .. (-2.5,0.6);
                    \draw[red, thick, snake it] (2.5,-1) .. controls (3,-2.1) and (4,-2.1) .. (4.5,-1);
                        
                 \end{tikzpicture}
             \end{scriptsize}
      
        \caption{The brane setup described in \cite{Armoni:2015jsa} for the case of $k \ge N_f/2$. The low-energy theory in the $p'$ vacuum is realised on the world-volume of the left hand stack of D3-branes, with the low-energy theory in the $p$ vacuum realised on the right stack. Modes living on the wall arise from the blue strings while the red strings are the bulk degrees of freedom. Note that we have branes (rather than anti-branes) on either side of the domain wall - this situation changes in the $k < N_f/2$ case and we expect a stack of branes on one side, and anti-branes on the other (so the arrows come with opposite orientation).}
        \label{fig: brane setup}
    \end{figure}

In some instances, the properties of domain walls indicate that they should be identified with D-branes in string theory. In particular, Witten demonstrated that the domain walls of pure ${\cal N}=1$ super Yang-Mills have a realisation in terms of type IIA D-branes \cite{Witten:1997ep}: they have the same tension and, moreover, the QCD-string can end on the domain wall in the same way the fundamental string can end on a D-brane.

We would like to suggest that the domain walls of 3d QCD can also be realised as D-branes.

The 2d field theory that lives on the domain walls, a gauged WZW model, can be deduced from anomaly cancellation arguments \cite{Armoni:2015jsa}. Moreover, the domain wall itself could be 'engineered' using a brane configuration, where the wall is a D-brane. The D-brane sits at a junction where on its one side 'lives' one of the 3d vacua and on its other side a different vacuum. 

Another realisation of the domain walls as D-branes is via holography. Let us use the holographic proposal in \cite{argurio_2020_vacuum}, focusing on $k \ge N_f/2$ (the case $k<N_f/2$ can be treated  similarly). Each vacuum of the model has a realisation in terms of a distribution of D-branes. The configuration consists of one stack of $p$ D-branes, another stack of $N_f-p$ branes and in addition a stack of Chern-Simons branes. The field theory on the Chern-Simons branes is a $U(\tilde N)_{N}$ theory - the same as the bosonised theory. Thus two different vacua of 3d QCD can be realised as two distributions of D-branes.

The wall tension is $N$ exactly as expected from a D-brane. Moreover, as in brane dynamics, where at $g_{\rm st}=0$ D-branes do not interact, in field theory domain walls do not interact and hence fundamental walls do not form a bound state: the composite walls 'disintegrates' into its constituents. At infinite $N$ it is possible to separate the fundamental walls and the $U(p'-p)$ symmetry is reduced to its Cartan, $U(1)^{(p'-p)}$ symmetry. At finite $N$ the walls attract and form a bound state, in agreement with D-brane attraction at non-zero $g_{st}$.

\section{Conclusions} 
\label{sec: conc}

In this paper we discussed various aspects of the domain walls that interpolate between the vacua of large-$N$ QCD\textsubscript{3}. We focused on the profile of the wall and the field theory that lives on the wall. We also commented on the similarity between the domain walls dynamics and D-branes dynamics. 

There are several interesting questions that we postpone for further studies: can we understand the domain wall as a single D-brane in a holographic setup? 

Interestingly, the gauged WZW model that lives on the domain wall is an interpolation between two configurations of Chern-Simons branes. This suggests that the domain wall theory could be obtained by placing two Chern-Simons branes on top of each other, one with $\tilde N_1$ branes and the other with $\tilde N_2$ anti-branes. We conjecture that the resulting theory on the Chern-Simons brane anti-brane system, after dimensional reduction along the $x$ direction, is a level $N$ gauged WZW model. After brane annihilation the remaining massless degrees of freedom on the wall correspond to level N WZW of a group of rank $p'-p$, with a global symmetry $U(p'-p)$.

Other questions include: what are the dynamics of $SO/Sp$ QCD theories and what are the walls of those theories? What is the dynamics of the walls at finite $N$? We expect that since at finite $N$ there is a single vacuum the wall should disintegrate in a Coleman De Luccia like mechanism \cite{Coleman:1980aw}.
We postpone the answers to those questions for future studies.

\vskip 1cm

{\Large \bf{Acknowledgments}} We thank Carlos Nunez for collaboration in initial stages of research. We thank Zohar Komargodski, Vasilis Niarchos and Shigeki Sugimoto for useful discussions. JW would also like to thank Dimitrios Chatzis for many helpful discussions. JW is supported by the STFC grant no. ST/Y509644/1

\appendix
\section{Derivation of the gauged WZW action}\label{sec: appendix}

    In this appendix we will start from the action \eqref{eqn: WZWs plus terms} plus the remaining bulk terms, and show that they can be rewritten as a gauged WZW model. For reference, the action \eqref{eqn: WZWs plus terms} is:

    \begin{align}
        S_{2d} = & \ S_{WZW}[H]_N - S_{WZW}[h]_N    \nonumber \\  & + N \int d^2x \ \frac{1}{8\pi} \mathrm{tr} \Big[ (B_mB^m - b_m b^m   ) - 2i (B_m \partial^m H H^\dagger - b_m \partial^m h h^\dagger)  \Big] \nonumber \\ & - N \int d^2 x \ \frac{i}{4\pi} \epsilon^{mn} \mathrm{tr} \Big[ \partial_m H H^\dagger B_n - \partial_m h h^\dagger b_n \Big]
    \end{align}

    and it comes with the following bulk terms:

    \begin{equation}
        S_{3d} =  \ N \int_{x>-X} d^3x \ \frac{1}{4\pi} \epsilon^{mxn} \ \mathrm{tr} \Big[ B_m \partial_x B_n - B_x F_{mn} \Big]  -   \frac{1}{4\pi} \epsilon^{mxn} \ \mathrm{tr} \Big[ b_m \partial_x b_n - b_x f_{mn} \Big].
    \end{equation}

    We can see that the bulk action contains $B_x$ and $b_x$ as Lagrange multipliers which have $F_{mn}=0$ and $f_{mn}=0$ respectively as their equations of motion. The Euler-Lagrange equations are insensitive to boundary terms, so in addition to the bulk equations of motion, in order for the variation of the action to vanish on the boundary we also require $B_x - i \partial_x H H^\dagger = 0$ and $b_x - i \partial_x h h^\dagger =0$ on the boundary.

    The bulk equations of motion for $B_x$ and $b_x$ have pure gauge solutions for $B_m$ and $b_m$, which means that we can write them as:

    \begin{equation}\label{eqn: pure gauge}
        B_m = -i U^\dagger \partial_m U, \qq b_m = -i V^\dagger \partial_m V \qq \mathrm{where} \qq U \in U(k+p'-N_f/2), \qq V \in U(k+p-N_f/2)
    \end{equation}

    and $U$ and $V$ transform in the fundamental representations of their respective groups. After doing this, the action becomes:

    \begin{align}
        S_{2d} = & \ S_{WZW}[H]_N - S_{WZW}[h]_N   \nonumber \\  & + N \int d^2x \ \frac{1}{8\pi} \mathrm{tr} \Big[ V^\dagger \partial_m V V^\dagger \partial^m V -  U^\dagger \partial_m U U^\dagger \partial^m U  \Big] \nonumber \\ & - N  \int d^2 x \ \frac{1}{4\pi}  \mathrm{tr} \Big[ U^\dagger \partial_m U \partial^m H H^\dagger -  V^\dagger \partial_m V \partial^m h h^\dagger + \epsilon^{mn}( \partial_m H H^\dagger U^\dagger \partial_n U - \partial_m h h^\dagger V^\dagger \partial_n V ) \Big]
    \end{align}

    with the bulk part

    \begin{equation}
        S_{3d} =  N \int_{x>-X} d^3x \ \frac{1}{12\pi} \epsilon^{\mu\nu\rho} \ \mathrm{tr} \Big[ U^\dagger \partial_\mu U U^\dagger \partial_\nu U U^\dagger \partial_\rho U - V^\dagger \partial_\mu V V^\dagger \partial_\nu V V^\dagger \partial_\rho V  \Big]
    \end{equation}

    so we are able to identify two more WZW actions, for $U$ and $V$. Now we define light-cone coordinates $x_+ = (t+y)/2$ and $x_- = (t-y)/2$ in which the action can be written:
    
    \begin{align}
        S_{2d} = & \  S_{WZW}[H]_N + S_{WZW}[U]_N - S_{WZW}[h]_N - S_{WZW}[V]_N \nonumber \\ & - N \int d^2x \frac{1}{4\pi} \mathrm{tr} \Big[ U^\dagger \partial_+ U \partial_- H H^\dagger - V^\dagger \partial_+ V \partial_- h h^\dagger \Big]
    \end{align}

    and now we are able to apply twice the Polyakov-Weigmann identity. Specifically, we will use:

    \begin{align}
        & S_{WZW}[U H]_N = S_{WZW}[U]_N + S_{WZW}[H]_N - \frac{N}{4\pi} \int d^2 x\ \mathrm{tr} \Big[ U^\dagger \partial_+ U  \partial_- H H^\dagger \Big] \\ & S_{WZW}[V h]_N = S_{WZW}[V]_N + S_{WZW}[h]_N - \frac{N}{4\pi} \int d^2 x \ \mathrm{tr} \Big[ V^\dagger \partial_+ V  \partial_- h h^\dagger \Big]
    \end{align}

    so that the action can now be written

    \begin{equation}
        S_{2d} = S_{WZW}[U H]_N - S_{WZW}[Vh]_N 
    \end{equation}

    and there is no bulk part (the bulk part that we started from has become the cubic part of the WZW actions after applying the solutions in \eqref{eqn: pure gauge}, which means that it should be thought of instead as part of the boundary action).

    The final step is to apply once more the Polyakov-Weigmann identity. If we make the following definitions:

    \begin{equation}
        Vh = f^\dagger \tilde{f} \qq \mathrm{and} \qq U H = f^\dagger H \tilde{f} \qq \mathrm{where} \qq f, \ \tilde{f} \in U(k+p-N_f/2)
    \end{equation}
    
    then as in \cite{Karabali:1989dk} the Polyakov-Weigmann identity allows us to write the action as:

    \begin{align}
        S_{2d} = S_{WZW}[U H]_N - S_{WZW}[Vh]_N   & = S_{WZW}[f^\dagger H \tilde{f}]_N - S_{WZW}[f^\dagger \tilde{f}]_N    \nonumber \\ & =   S_{gWZW}[H, A]_N
    \end{align}

    with the components of the gauge field $A$ given by $A_+ = \partial_+ f f^\dagger$ and $A_- = \partial_- \tilde{f} \tilde{f}^\dagger$ so $A$ is a gauge field for $U(k+p-N_f/2)$. We are left with a gauged WZW action, which can be written as

    \begin{equation}
        S_{gWZW}[H,A]_N = S_{WZW}[H]_N + \frac{N}{4\pi} \int d^2 x \ \mathrm{tr}\Big[ A_+ \partial_- H H^\dagger - A_- H^\dagger \partial_+ H + A_+ H A_- H^\dagger - A_- A_+ \Big]
    \end{equation}

    as given in \cite{Karabali:1989dk}.

\newpage

\providecommand{\href}[2]{#2}\begingroup\raggedright\endgroup

\end{document}